\documentclass[11pt]{article}
\usepackage[utf8]{inputenc}
\usepackage[english]{babel}
\usepackage{subcaption} 
\usepackage{graphicx}
\usepackage{amsmath}
\usepackage{upgreek}
\usepackage{amssymb}
\usepackage[unicode]{hyperref}
\usepackage[tableposition=top]{caption}
\usepackage{multirow}
\usepackage{tabularx}
\usepackage{textcomp}
\usepackage{gensymb}
\usepackage{graphics}
\usepackage{color}
\usepackage[dvipsnames]{xcolor}
\usepackage[a4paper, total={6.5in, 8.5in}]{geometry}
\usepackage{tikz}
\usetikzlibrary{backgrounds}

\numberwithin{figure}{section}
\numberwithin{table}{section}
\usepackage{chngcntr}
\counterwithout{figure}{section}
\counterwithout{table}{section}
\usepackage{soul}
\usepackage{gensymb}

\usepackage{fancyhdr}
\usepackage{tikz}
\usetikzlibrary{backgrounds}

\title{A low cost, flexible atmospheric pressure plasma jet device with good antimicrobial efficiency}

\begin{document}

\pagestyle{fancy}
\fancyhead{} 
\fancyhead[L]{\textbf{\tiny This article has been accepted for publication in IEEE Transactions on Radiation and Plasma Medical Sciences. This is the author's version which has not been fully edited and content may be different from the final publication. Citation information: DOI \href{https://dx.doi.org/10.1109/TRPMS.2023.3342709}{10.1109/TRPMS.2023.3342709}}}

\author{Fellype~do~Nascimento\textsuperscript{1}, Aline~da~Gra\c{c}a~Sampaio\textsuperscript{2}, Noala~Vicensoto~Moreira~Milhan\textsuperscript{2},\\Aline~Vidal~Lacerda~Gontijo\textsuperscript{3}, Philipp~Mattern\textsuperscript{4,5}, Torsten~Gerling\textsuperscript{4,5},\\Eric~Robert\textsuperscript{6}, Cristiane~Yumi~Koga-Ito\textsuperscript{2}, Konstantin~Georgiev~Kostov\textsuperscript{1}\\ \small{\textit{\textsuperscript{1} Faculty of Engineering in Guaratinguet\'{a}, UNESP, Guaratinguet\'{a}, SP, Brazil}}\\ \small{\textit{\textsuperscript{2} Institute of Science and Technology, UNESP, S\~{a}o Jos\'{e} dos Campos, Brazil}}\\ \small{\textit{\textsuperscript{3} Department of Pharmacy, Anhanguera University, S\~{a}o Jos\'{e} dos Campos, Brazil}}\\ \small{\textit{\textsuperscript{4} ZIK plasmatis, Leibniz Institute for Plasma Science and Technology (INP), Greifswald, Germany}}\\ \small{\textit{\parbox{15cm}{\centering \textsuperscript{5} Diabetes Competence Centre Karlsburg (KDK), Leibniz Institute for Plasma Science and Technology (INP), Karlsburg, Germany}}}\\ \small{\textit{\textsuperscript{6} GREMI, CNRS/Universit\'{e} d'Orl\'{e}ans, Orl\'{e}ans, France}}
}

\maketitle

\begin{abstract}
Plasma sources suitable to generate low temperature plasmas has been fundamental for the advances in plasma medicine. In this research field, plasma sources must comply with stringent conditions for clinical applications. The main requirement to be met is the patient and operator's safety and the ethical requirement of effectivity, which encompasses the electrical regulations, potential device toxicity and effectiveness in relation to the desired treatment. All these issues are addressed by the German pre-standard DIN SPEC 91315:2014-06 (DINSpec), which deals with the safety limits, risk assessment and biological efficacy of plasma sources aimed for medical applications. In this work, a low cost, user-friendly and flexible atmospheric pressure plasma jet (APPJ) device was characterized following the DINSpec guidelines. The device, which is still under development, proved to be safe for medical applications. It is capable of producing an APPJ with low patient leakage current and UV emission, gas temperature lower than 40 {\textdegree}C, production of harmful gases within the safety limits and low cytotoxicity. The most differentiating feature is that the device presented good antimicrobial efficacy even operating at frequency of the order of just a few hundred Hz, a value below that of most devices reported in the literature.
\end{abstract}

\\
\\
\tikzstyle{background rectangle}=[thin,draw=black]
\begin{tikzpicture}[show background rectangle]

\node[align=justify, text width=0.9*\textwidth, inner sep=1em]{
{\small This article has been accepted for publication in IEEE Transactions on Radiation and Plasma Medical Sciences. This is the author's version which has not been fully edited and content may be different from the final publication. Citation information: DOI \href{https://dx.doi.org/10.1109/TRPMS.2023.3342709}{10.1109/TRPMS.2023.3342709}}
};

\node[xshift=3ex, yshift=-0.7ex, overlay, fill=white, draw=white, above
right] at (current bounding box.north west) {
\textit{Dear reader,}
};

\end{tikzpicture}

\section{Introduction}
During the last two decades plasma sources aimed to produce atmospheric pressure plasmas (APPs) and their applications have received a lot of attention. Since then, APPs have been used for the treatment of materials, for medical and biomedical applications \cite{dubuc_use_2018, duarte_comprehensive_2020, borges_applications_2021, adamovich_2022_2022, deepak_review_2022, das_role_2022}. Medical and biomedical applications, including dentistry, that employ cold atmospheric pressure plasmas (CAPPs) have advanced significantly and are becoming therapeutic alternatives or part of combination treatment for some conventional treatments \cite{bekeschus_white_2019, von_woedtke_perspectives_2020, liu_cold_2020, dubey_cold_2022, zhou_cold_2022, chupradit_recent_2023, matthes_efficiency_2022, matthes_-vitro_2023}. This is happening due to the recent advances in the development of plasma sources that produce cold plasmas with gas temperature close to that of the environment, combined with the generation of a large amount of reactive species among other relevant properties \cite{liu_cold_2020, laroussi_cold_2020, bekeschus_medical_2021, miebach_medical_2022, berner_gas_2023}.

The treatment of materials or tissues using APPs can be done in basically two ways. One of them is with the plasma impinging/touching the target, which will be called direct treatment, and the other way is with the plasma without any contact with the target (indirect treatment). In a direct treatment the action of the plasma on the material is a combination of physical and chemical processes \cite{laroussi_plasma_2018, lin_map_2021, ma_plasma-controlled_2022} and is subjected to the type of treated material \cite{sobota_plasma-surface_2019, teschner_investigation_2019}. The physical action occurs mainly due to the electric current, electric field, the gas temperature and the gas flow velocity \cite{zhang_synergistic_2014, vijayarangan_cold_2018}. The chemical action is mainly due to the interaction of the reactive species produced by the plasma with the substrate \cite{lu_reactive_2016, khlyustova_important_2019, busco_emerging_2020}. In indirect treatments only the reactive species are supposed to interact with the target. Of course, in situations in which the plasma source produces a plasma jet, the gas flow can also produce some physical effect on the target due to the gas temperature, even if the plasma does not touch the target \cite{tanaka_state_2017, saadati_comparison_2018, brany_cold_2020, terefinko_biological_2021}. In addition, ultraviolet (UV) radiation can be produced by the plasma source in both direct and indirect treatments \cite{hahn_concept_2020, miebach_conductivity_2022}. Plasma activated liquids (PAL) also performs indirect plasma treatments on the substrates in contact with the liquids \cite{thirumdas_plasma_2018, mai-prochnow_interactions_2021, tornin_evaluation_2021, milhan_applications_2022}. The PAL action on the substrate is exclusively due to the presence of reactive species in the liquid.

In medical and biomedical applications, some processes are favored when using a direct plasma treatment due to the higher degree of interaction between the plasma and the tissue \cite{miebach_conductivity_2022, he_efficacy_2020, kenari_therapeutic_2021, mrochen_toxicity_2022}. However, direct plasma treatments may present some risks to the target tissue, since in this operating mode the plasma has physical properties that can cause damage if the plasma device is not correctly set up, and such damage can be serious for human tissues if above a critical safety level \cite{weltmann_atmospheric_2009, gerling_relevant_2018, najafzadehvarzi_risk_2022}. Electrical current and gas temperature are the main discharge properties to be considered when performing a direct treatment on human tissues, being that the first must not exceed 100 {\textmu}A-rms in AC mode or 10 {\textmu}A in DC operation and the second must not exceed 40 {\textdegree}C \cite{noauthor_iec_2015, mann_introduction_2016}. Regarding electrical safety, the situation with the accumulation of short duration pulses, as delivered by many APP sources, is so far not so well-documented and regulated. In addition to the safety related to the discharge properties, UV radiation and toxic gasses produced within the plasmas lead to limitations of exposure time to plasma treatment \cite{gerling_relevant_2018, mann_introduction_2016, timmermann_piezoelectric-driven_2020}. The UV radiation directly affects human tissues and the maximum daily exposure ($D_{max}$) to UV radiation is limited to 3 mJ/cm${}^2$ \cite{protection_guidelines_2004}. The toxic gasses that can be produced by an APP, like ozone ($\rm{O_3}$), nitrite ($\rm{NO_2}$) and nitrate ($\rm{NO_3}$), do not cause damage to the target tissue, but can affect human lungs if their concentration in air is above the safe levels. When working in an environment for 8 hours, the maximum recommended concentration of $\rm{O_3}$ in air is 0.055 ppm and accumulation in a treatment room needs to be considered \cite{hilker_use_2018, noauthor_directive_2002}. For $\rm{NO_2}$ in the same working conditions the maximum recommended value is 0.019 ppm \cite{noauthor_directive_2008}.

To comply with these limitations, any plasma source for medical applications should be characterized to ensure that the above mentioned quantities are within the safety levels. In this way, the DIN SPEC 91315:2014–06 (to be referred as DINSpec from now on) was established, which is a German pre-standard based in part on DIN EN 60601-1, which relates to IEC 60601-1. The DINSpec deals with the safety limits of plasma sources for medical applications and suggests a protocol for risk assessment of plasma sources and establishes basic parameters for checking the biological efficacy of a device \cite{mann_introduction_2016, timmermann_piezoelectric-driven_2020, noauthor_din_2014}. The DINSpec protocol can be applied to any plasma source. However, some parameters like the patient leakage current (PLC) and gas temperature ($T_{gas}$) are more important when using devices that produce plasma jets since, in this situation, such parameters can increase with the distance from target, instead of only decreasing as with devices that do not employ a gas flow or are aimed to be used for indirect treatment \cite{miebach_conductivity_2022, najafzadehvarzi_risk_2022, nascimento_gas_2023}.

Although low frequency plasma sources are common for material treatment, in general, the ones aimed to be used in medicine and dentistry operate at frequencies higher than the ones in the electrical line (50/60 Hz) \cite{khlyustova_important_2019, kim_review_2023, reuter_kinpenreview_2018}. Some of them use radio frequency (RF) power supplies to generate the plasmas. Others employ power supplies with frequencies ranging from 200 Hz to a few tens of kHz \cite{duske_comparative_2015, hertel_antibacterial_2018, chaerony_siffa_development_2022}. In the opposite direction, Xaubet \textit{et al} reported the operation of the Magiplas device whose voltage waveform is a 50 Hz sinusoidal burst modulated at 2 Hz, which can be seen as a pulsed signal with a frequency of 2 Hz \cite{xaubet_design_2018}. Even using a low frequency plasma source, a good antimicrobial efficacy was achieved with such equipment.

The operating frequency of a power supply used to ignite an APP has direct influence on the discharge parameters and behavior \cite{kim_review_2023, tendero_atmospheric_2006, winter_atmospheric_2015}. One of the effects is that the discharge ignition tends to be easier at higher frequencies, which allows a reduction in the voltage amplitude required for the ignition. Another one is related to the power dissipated on a discharge ($P_{dis}$), which is directly proportional to the voltage frequency \cite{gerling_power_2017}. This last fact has a direct influence on the gas temperature of an APP, with higher $P_{dis}$ usually leading to higher $T_{gas}$ values. An advantage of plasma sources that operate at higher frequencies is that the effective plasma duration is higher than in those devices operating at lower frequencies. In addition, the production of reactive species can also be enhanced by operation at higher frequencies \cite{yang_variable_2013, baek_effects_2016}. Those facts can lead to higher efficiency in the plasma treatment with shorter exposure time. However, the plasma source must be very well designed in this case in order to avoid the excessive heating related to the high frequency operation.

In this work, the results of characterization and antimicrobial efficiency of a low cost and low frequency APPJ device aimed for applications in dentistry and medicine are presented. The device characterization aimed to check if the device complies with safety requirements for plasma sources to be applied as a medical tool. Such characterization was accessed by carrying out a set of measurements suggested by the DINSpec. The biological assays with different sets of bacteria and fungi, as proposed by the DINSpec, were also carried out in order to ensure that the plasma jet produced with the device has antimicrobial properties. Experiments were also performed in order to ensure that the APPJ has no cytotoxic effects.

\section{Materials and methods}
\subsection{Characterization of the plasma sources for medical applications}
The characterization of the plasma source used in this work was carried out performing a set of measurements suggested by the DINSpec, which consists mainly of:
\begin{itemize}
 \item ensure the rms values of electrical current reaching the patient (PLC as abbreviation for the so-called patient leakage current) $\rightarrow$ must not exceed 100 {\textmu}A-AC
 \item determine the effective UV radiation emitted by the plasma $\rightarrow$ will define the maximum exposure time to plasma treatment
 \item measure the temperature of the gas ($T_{gas}$) and the target ($T_{targ}$, a copper plate) $\rightarrow$ it is not recommended that the maximum value of $T_{gas}$ exceeds 40 {\textdegree}C and the maximum value of $T_{targ}$ must be less than 40 {\textdegree}C
 \item the generation of ozone ($\rm{O_3}$), nitrogen oxide ($\rm{NO}$) and nitrite ($\rm{NO_2}$) gases produced by the plasma $\rightarrow$ define the maximum exposure time to plasma treatment and minimum distance from the plasma outlet
 \item check for antimicrobial efficacy, cellular toxicity, etc.
\end{itemize}

\subsection{Portable plasma device and characterization processes}
\begin{figure}[t]
\centering
\begin{subfigure}{0.48\textwidth}
\centering
\includegraphics[width=7.5 cm]{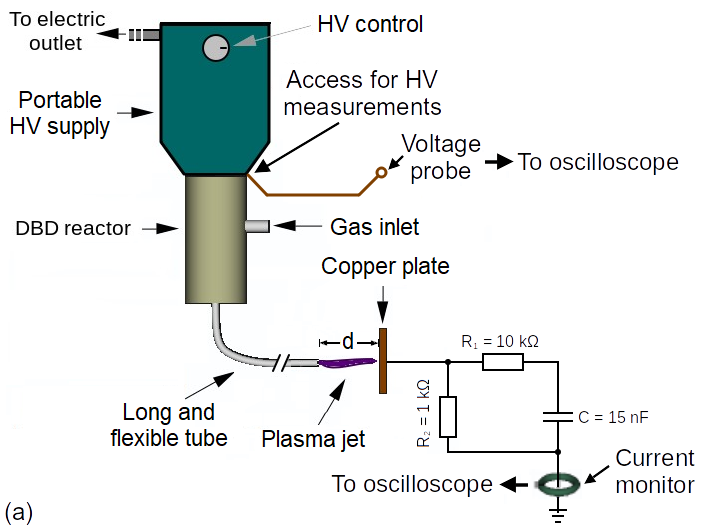}
\end{subfigure}
\begin{subfigure}{0.43\textwidth}
\centering
\includegraphics[width=5.28 cm]{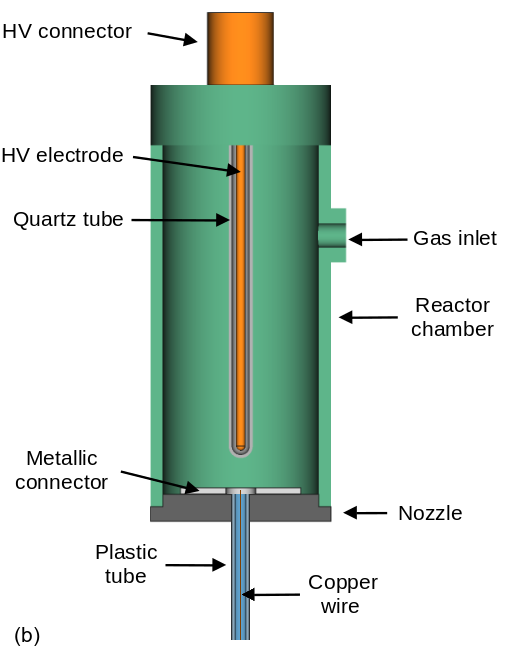}
\end{subfigure}
\caption{Portable plasma source scheme. (a) Overview of the plasma source. (b) DBD reactor in detail. Elements are out of scale.\label{plSrcSch}}
\end{figure}

Figure~\ref{plSrcSch}(a) shows an overview of the plasma source and the main scheme for obtaining the waveforms of applied voltage ($V$), discharge current ($i$) and PLC when producing a plasma jet impinging on a copper (Cu) target. Figure~\ref{plSrcSch}(b) presents a detailed view of the dielectric barrier discharge (DBD) reactor which integrates the plasma source.
The plasma source used in this work consists of a portable power supply and a DBD reactor to which is connected a 1.0 m long flexible plastic tube, whose outer and inner diameters are 4.0 mm and 2.0 mm, respectively. The DBD reactor is composed of a dielectric enclosure (with inner diameter of 10 mm) containing a pin-electrode (1.8 mm in diameter) made of tungsten encapsulated by a closed-end quartz tube with outer and inner diameters equal to 4.0 mm and 2.0 mm, respectively. The pin-electrode is in turn connected to a male metallic socket, which is attached to the dielectric enclosure and plugged into the female socket of the power supply. Inside the long tube there is a thin copper wire (0.5 mm in diameter), which is fixed to a metallic connector placed inside the reactor chamber. The portable power supply was adapted from a commercial device used in aesthetic applications. It is capable of generating a sequence of damped-sine waveforms with peak voltage values up to 20 kV and oscillating frequency of nearly 110 kHz. Each damped-sine waveform behaves as a high voltage (HV) burst signal (see Fig. 1(b) of \cite{nascimento_gas_2023} for a detailed view of the HV waveform). Although the voltage oscillation inside each burst is relatively high, the power supply is able to produce no more than four pulses within each 50 or 60 Hz cycle, depending on the electrical network frequency.

The basic operating principle of the plasma jet device is as follows: when the reactor chamber is fed with a working gas (He in this work) and the HV is switched on, a primary DBD discharge is ignited inside the reactor. So, the plasma impinges on the metallic connector, which is not in touch with the quartz tube and acts as a floating electrode. As a result, the thin Cu wire inside the plastic tube acquires electric charge creating an intense electric field at the wire tip. Thus, when the He gas flowing through the long plastic tube reaches its distal end a secondary discharge is ignited, producing a plasma jet.

\begin{figure}[t]
\centering
\begin{subfigure}{0.48\textwidth}
\includegraphics[width=8. cm]{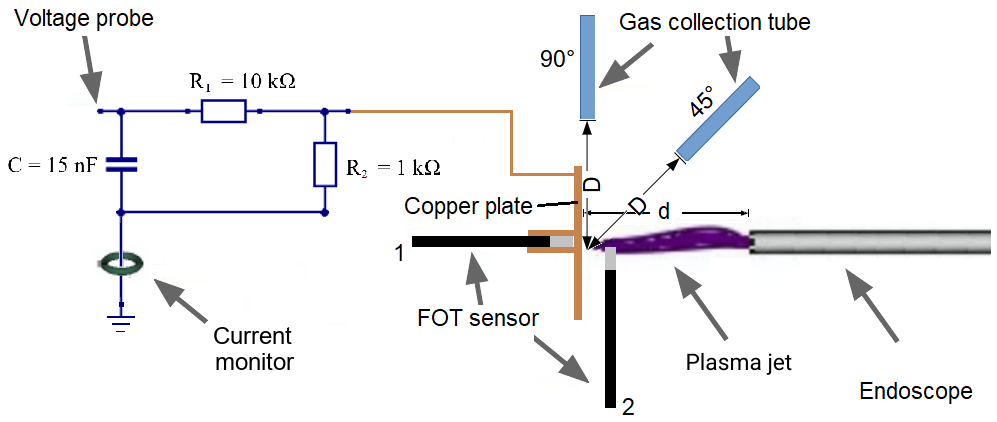}
\caption{{}}
\end{subfigure}
\begin{subfigure}{0.48\textwidth}
\includegraphics[width=7.5 cm]{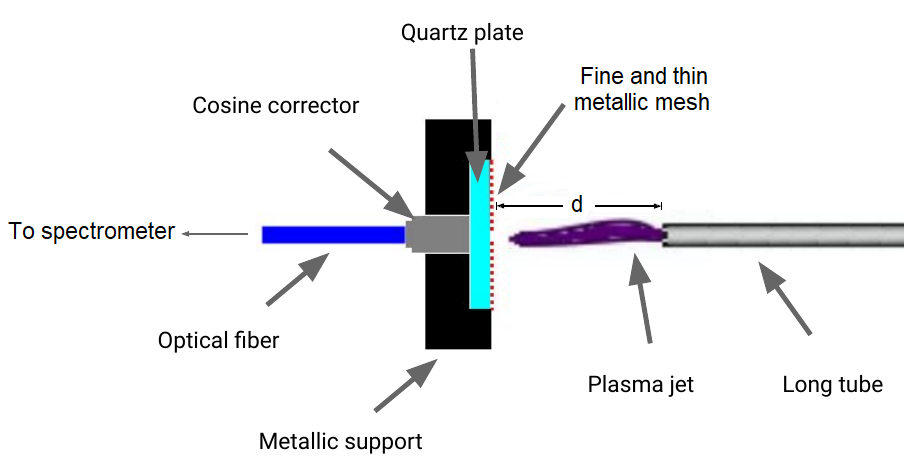}
\caption{{}}
\end{subfigure}
\caption{Experimental setup for the characterization of the plasma source. (a) Scheme for measurements of PLC, temperatures and production of harmful gasses. (b) Scheme used for OES measurements.\label{expSetPlSrc}}
\end{figure}

The experimental setup used in the characterization of the plasma source is shown in Fig.~\ref{expSetPlSrc}. Figure~\ref{expSetPlSrc}(a) shows the setup used for measurements of gas and target temperatures, production of $\rm{O_3}$ and $\rm{NO_2}$ as well as a more detailed view of the PLC measurement scheme, while the details of the OES measurements are depicted in Fig.~\ref{expSetPlSrc}(b).

The measurements of gas and Cu plate temperatures ($T_{gas}$ and $T_{Cu}$, respectively) were carried out simultaneously employing two fiber optic temperature (FOT) sensors, from LumaSense Technologies Inc. GmbH, USA (model FOT Lab Kit). The data acquisition for gas temperature measurements were carried out as a function of the distance $d$ from the plastic tube outlet, keeping the gas flow rate at 2.0 slm. For the temperature measurements as a function of $d$, the long tube outlet remained at a fixed position and the $d$ values were changed using a horizontal displacer to which the Cu plate and FOT sensors were attached. When measuring $T_{Cu}$ and $T_{gas}$ the distance between the Cu plate and the FOT 2 was kept unchanged and equal to 1.0 mm. The Cu plate is a square piece 1 cm sided and 0.09 mm in thickness, with a mass of 0.312 g. The FOT sensors allow temperature measurements as a function of time. Thus, both $T_{gas}$ and $T_{Cu}$ were measured for {$\sim$}50 s with the power supply turned off, without discharge ignition, and for {$\sim$}70 s with the power supply turned on, with discharge ignition and producing a plasma jet. The time averaged values of $T_{gas}$ and $T_{Cu}$ with discharge ignition were then calculated and used as the gas and target temperatures for the plasma on phase.

The electrical characterizations of the device consisted mainly of obtaining the discharge power ($P_{dis}$) and patient leakage current (PLC) as a function of $d$. For this purpose, the applied voltage ($V(t)$) was measured with a 1000:1 voltage probe and the electrical current ($i(t)$) that flows through the entire circuit was measured with a current monitor placed after the RC circuit shown in Fig.~\ref{expSetPlSrc}. All the electrical signals were recorded with an oscilloscope and the $P_{dis}$ values were calculated using Eq.~\ref{eqpower}, where $f$ = 50 Hz is the repetition rate of the voltage signal, with $T = f^{-1}$.

\begin{equation}
{P_{dis} = f \int _{0} ^{T} v(t) \cdot i(t) dt} \label{eqpower}
\end{equation}

PLC measurements were performed with temporal resolution of the PLC waveform with the RC circuit shown in Fig.~\ref{expSetPlSrc}. So, the PLC values were calculated using the voltage $V_{C}(t)$ measured across the capacitor C by applying:
\begin{equation}
{PLC = \sqrt{\frac{1}{T} \int _{t_i} ^{t_f} \left( {\frac{V_{C}(t)}{R_2}} \right) ^2(t) dt}} \label{eqPLC}
\end{equation}

\noindent that is, the PLC value is the effective (root mean square – RMS) value of the $V_{C}(t)/R_2$ signal.

Optical emission spectroscopy (OES) was employed for identification of emitting species and for calculations of the effective UV radiation emitted ($E_{eff}$) by the plasma jet. All the OES measurements were performed with the plasma jet impinging on a metallic mesh over a quartz plate. The measurement setup is depicted in Fig.~\ref{expSetPlSrc}(b). A spectrometer (Avantes, AvaSpec 3648, 200–1100 nm range, {$\sim$}2 nm resolution, 300 lines/mm grating, blazed at 300 nm, deep-UV-detector coated CCD linear array) with an optical fiber connected to a cosine corrector was employed for OES measurements. The OES system was calibrated for absolute light intensity measurements. The distance $d$ between the plasma outlet and target was varied from 4 mm to 40 mm.

One of the most important assessments to be done through OES measurements is the determination of ($E_{eff}$) values, which are calculated for each $d$ value using:

\begin{equation}
{E_{eff} = \int _{\lambda_1} ^{\lambda_2} E(\lambda) \cdot s(\lambda) d{\lambda}} \label{eqEeff}
\end{equation}

\noindent with $\lambda_1$ = 200 nm and $\lambda_2$ = 400 nm; $s(\lambda)$ is the spectral weighting function~\cite{timmermann_piezoelectric-driven_2020}.

Those $E_{eff}$ values are then used to determine the maximum plasma exposure time ($t_{exp}$) due to UV radiation produced by the plasma jet, which is defined as:

\begin{equation}
{t_{exp} = \frac{D_{max}}{E_{eff}}}\label{eqtmax}
\end{equation}

\noindent where $D_{max}$ = 3 mJ/cm${}^2$ is the maximum UV daily dose that a human tissue can receive \cite{protection_guidelines_2004}.

The OES measurements were also used to calculate the vibrational temperature ($T_{vib}$) values of nitrogen molecules using the band emissions from the $\rm{N_{2}}$ second positive system ($C {}^{3} \Pi_u, \nu' \rightarrow B {}^{3} \Pi_g , \nu''$) with $\Delta \nu$ = $\nu' - \nu''$ = $-2$, in the wavelength range from 365 nm to 382 nm \cite{zhang_determination_2015, ono_optical_2016}. Simulations of the $\rm{N_{2}}$ emission spectra were performed using a software called massiveOES \cite{vorac_batch_2017, vorac_state-by-state_2017} and were compared to the experimental spectra. Then, the $T_{vib}$ values are those that generate simulated curves which best fit the experimental data.

In addition to the UV irradiation and determination of the vibrational temperature in the APPJ, the OES measurements were also used to estimate the intensity emissions of some RONS (namely $\rm{NO}$, $\rm{OH}$ and $\rm{O}$) as a function of the distance to the plasma outlet.

The concentrations of $\rm{O_3}$ and $\rm{NO_2}$ gases produced within the plasma jet were measured using commercially available gas detectors from Horiba (models APOA-360 for $\rm{O_3}$ and APNA-370 for $\rm{NO_2}$). The scheme for such measurements is depicted in Figure~\ref{expSetPlSrc}(a). For both gasses, the measurements were carried out with the gas collection tube placed at angles of 45{\textdegree} and 90{\textdegree} in relation to the plasma jet axis. Then, measurement data were collected as a function of the distance $D$ between the spot where the plasma impinges on the Cu plate and the inlet of the gas collection tube. The distance $d$ between plasma outlet and target was fixed at 15.0 mm. Additional measurements in the free jet condition were carried out for evaluation of the $\rm{O_3}$ production. In this case, the collecting tube was placed in front of the plasma jet, that is, at an angle of 180{\textdegree} in relation to the jet axis and the data was collected as a function of $D$.

Image diagnostics of the gas flow were also carried out in this work. For such purpose, Schlieren measurements were performed for the He gas expanding with and without discharge ignition. This was done for a He flow rate of 2.0 slm with the gas being flushed in the horizontal direction. The helium flow at the capillary exit was visualized using a Schlieren optical bench assembling two plane-convex lenses having an aperture of 75 mm and a focal length of 150 mm (Thorlabs LA1002). The light source was a green light diode emitting at 530 nm (Thorlabs M530L4). A high frame rate CCD camera (Photron Fastcam V4) was used to capture Schlieren patterns. The images were acquired using a 500 fps frame rate and 1/20000 aperture (50 {\textmu}s exposure time).

\subsection{Antimicrobial assay}
A Gram positive bacterial species (\textit{Staphylococcus aureus}, ATCC 6538), a Gram negative species (\textit{Pseudomonas aeruginosa}, ATCC 15442) and a fungal species (\textit{Candida albicans}, ATCC 18804) were used for the assays. Fresh cultures of \textit{S. aureus}, \textit{P. aeruginosa} and \textit{C. albicans} were obtained by plating on Brain Heart Infusion (BHI) agar and Sabouraud agar, respectively. Plates were incubated at 37 {\textdegree}C for 24 h, under aerobiosis. To evaluate the antimicrobial effect of the plasma jet, standardized microbial suspensions (106 UFC/mL) were prepared in physiologic solution (NaCl 0.9{\%}) with the aid of a spectrophotometer. Afterwards, an aliquot of 100 {\textmu}L was plated on the surface of the culture medium with the aid of a sterile swab, followed by drying for 15 minutes in an aseptic environment. Then, the plasma jet was applied perpendicularly on Petri dishes. The microorganisms were exposed to different treatment conditions with variations of discharge power, distance from plasma outlet (15 and 20 mm) and exposure time (1.0, 2.0, 3.0 and 5.0 min). Each experiment was performed in triplicate at two different times. The control group was exposed to He gas without ignition (no plasma). The plates were incubated at 37 {\textdegree}C for 24 h and the diameters of the circular zone of the microbial inhibition were measured with a ruler and the area of microbial inhibition zones were calculated by the area of the circle.

\begin{figure}[!t]
\centering
\includegraphics[width=8.5 cm]{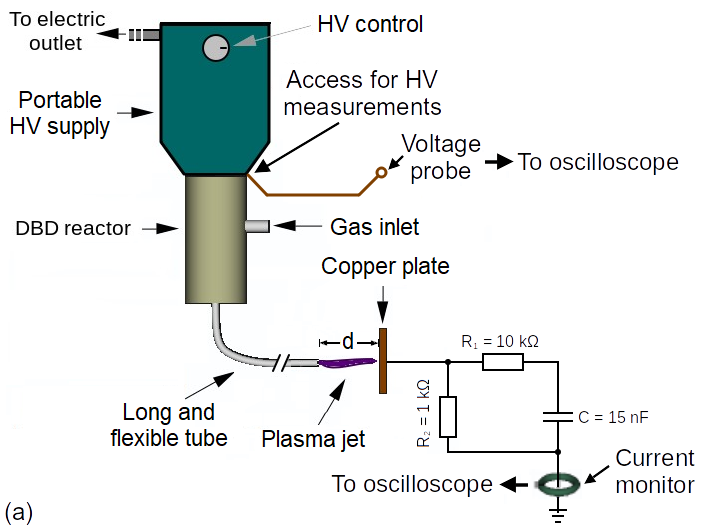}
\caption{Scheme used for exposure of microorganisms to the plasma jet.\label{bioSetup}}
\end{figure}

Since the biological assays employ an experimental setup in which the plasma jet interacts with agar substrate inside a Petri dish instead of directly with the copper plate like in Fig.~\ref{expSetPlSrc}, with a plastic holder between the Petri dish and the copper plate, changes in the $P_{dis}$ values were observed. Thus, new measurements for $P_{dis}$ taking into account the setup shown in Fig.~\ref{bioSetup} were carried out by measuring the input high voltage and the current that flows through the resistor R and then applying Eq.~\ref{eqpower}.

\subsection{Cytotoxicity evaluation}
The cell viability was evaluated after exposure of normal oral keratinocytes (NOK) and fibroblasts (3T3) to the APPJ. The cells were grown in Dulbecco's Modified Eagle's medium (DMEM), supplemented with 10{\%} of fetal bovine serum and 1{\%} of penicillin (100 U/mL)/streptomycin (100 mg/mL) and maintained at 37 {\textdegree}C and 5{\%} $\rm{CO_2}$. For the assay, the cells were seeded at a density of 8x10${}^3$ cells per well in 96-well plates and incubated for 24 h to allow the cell adhesion. Then, the cells were exposed to the most effective operating parameter of the plasma jet device, regarding the antimicrobial activity. The amount of 30 {\textmu}l of Hanks' Balanced Salt Solution (HBSS) was added to the wells to prevent their drying out during the plasma exposure. Afterwards, the cells were incubated at 37 {\textdegree}C for 24 h. For the measurement of cell viability, 100 {\textmu}L of 3-(4,5-Dimethylthiazol-2-yl)-2,5-diphenyl tetrazolium bromide (MTT) was added to the wells. After 1 h, formazan crystals were dissolved with Dimethyl Sulfoxide (DMSO). The resulting optical density of the solution was obtained in a spectrophotometer at 570 nm. Absorbance data were normalized to the untreated control group (= 100{\%}). Two independent experiments were carried out with six replicates each (n = 12). The cytotoxicity threshold was set at 70{\%} according to ISO 10993-5 \cite{international_organization_for_standardization_iso_2009}.

\section{Results and discussion}
This section is divided into two main subsections. In the first section the results from the characterization of the plasma source and in the second one the results from the biological assays are presented.

\subsection{Characterization of the plasma source}
\subsubsection{Discharge power and patient leakage current}
\begin{figure*}
\centering
\begin{tabular}{cc}
\begin{minipage}[t]{0.48\textwidth}
\centering
\includegraphics[width=6.8 cm]{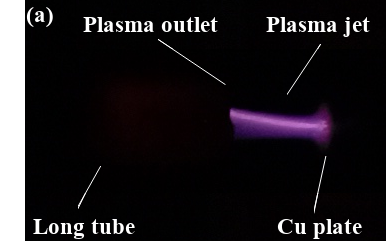}
\end{minipage}

&

\begin{minipage}[t]{0.48\textwidth}
\centering
\includegraphics[width=6.8 cm]{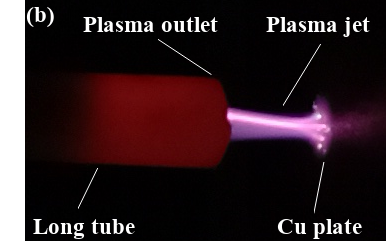}
\end{minipage}

\\

\begin{minipage}[t]{0.48\textwidth}
\centering
\includegraphics[width=7.5 cm]{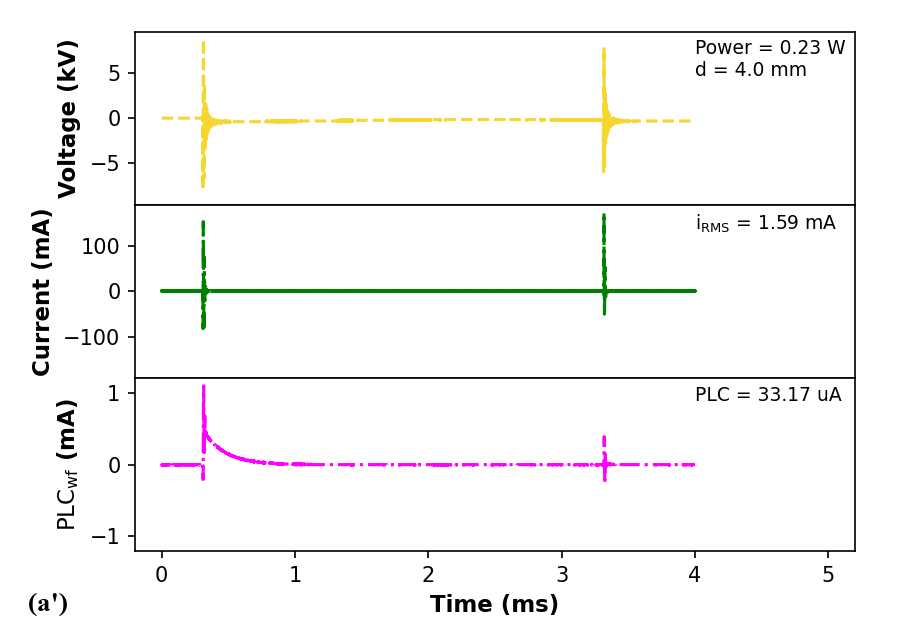}
\end{minipage}

&

\begin{minipage}[t]{0.48\textwidth}
\centering
\includegraphics[width=7.5 cm]{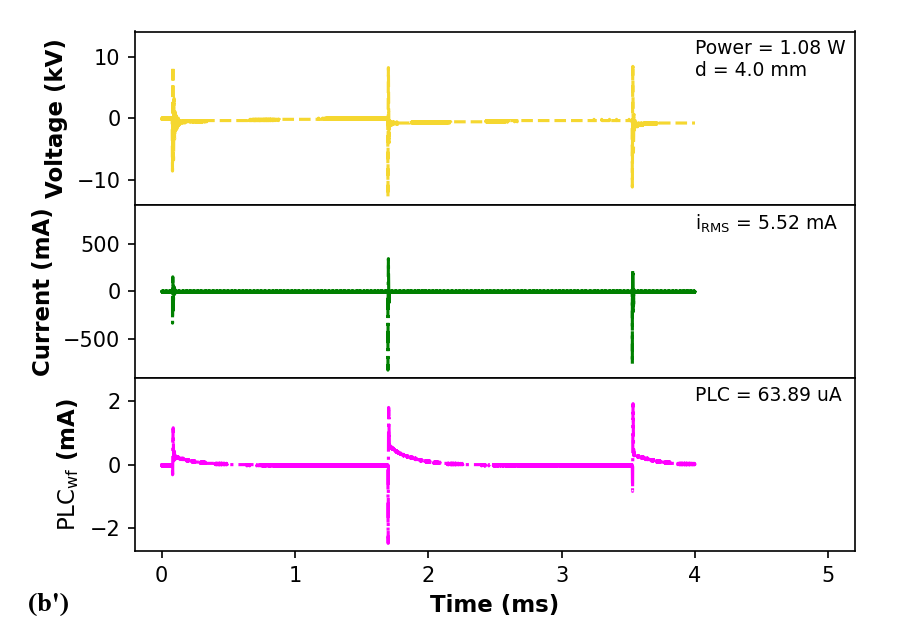}
\end{minipage}

\\

\begin{minipage}[t]{0.48\textwidth}
\centering
\includegraphics[width=7.5 cm]{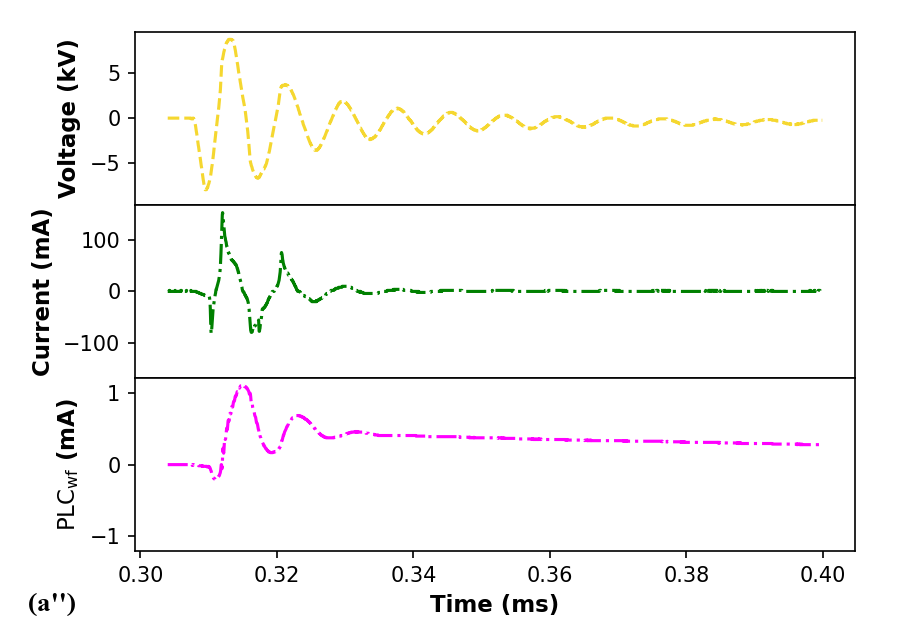}
\end{minipage}

&

\begin{minipage}[t]{0.48\textwidth}
\centering
\includegraphics[width=7.5 cm]{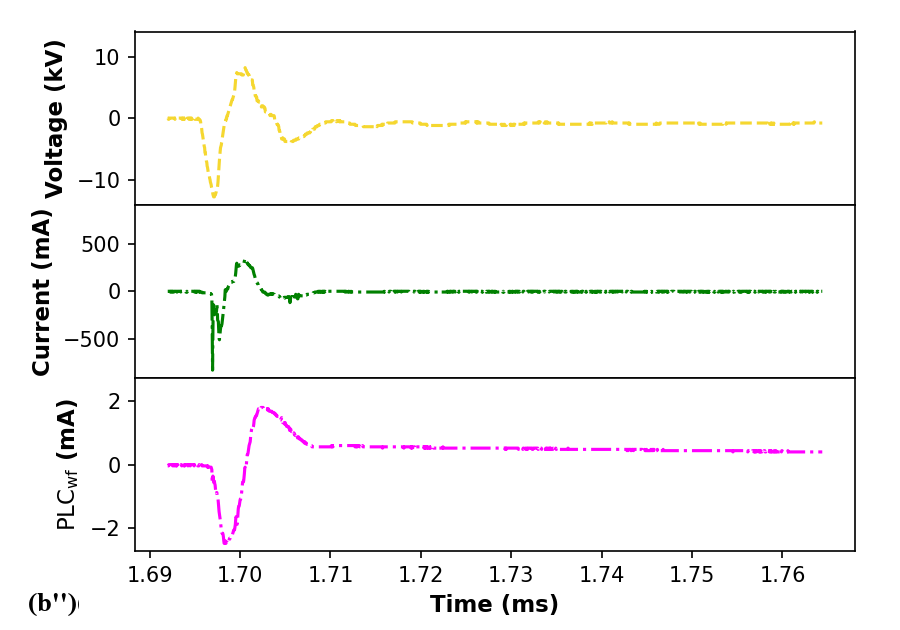}
\end{minipage}

\end{tabular}
\caption{Photos of the APPJ operating in (a) lower and (b) higher power conditions with their corresponding waveforms measured for voltage, current and $\rm{PLC_{wf}}$ (a' and b'). $d$ was equal to 4.0 mm in both cases. (a'') and (b'') show magnified views for the first and second pulses in the lower and higher power cases, respectively.}\label{wfEg}

\end{figure*}

In this study, the electrical characterizations of the plasma source were carried out for two working conditions, which differ in the intensity and number of pulses of the applied voltage. In the first case, the maximum peak-to-peak voltage ($V_{pp}$) is nearly 15 kV which generates plasma discharges with low dissipated power, with a maximum value of the order of 0.3 W. In the second case, the maximum $V_{pp}$ value is close to 30 kV and the maximum dissipated power is nearly 1.0 W. We will call the first and second cases of lower power and higher power conditions, respectively. Figure~\ref{wfEg} shows the discharge photos for (a) the lower and (b) higher power cases with the corresponding examples of voltage, current and PLC (labeled as $\rm{PLC_{wf}}$ in the figures) waveforms measured at a distance of 4.0 mm from the plasma outlet in (a') and (b'), respectively. The voltage and current waveforms were used to calculate the $P_{dis}$ values by applying equation~\ref{eqpower}. The $\rm{PLC_{wf}}$ curves were used to calculate the PLC values by applying equation~\ref{eqPLC}. The temporal evolution of the electrical signals were recorded for ten consecutive times for each distance from the plasma outlet. Such waveforms were used to calculate the respective values of $P_{dis}$ and PLC for each measurement. Then, both $P_{dis}$ and PLC values were averaged over ten measurements.

\begin{figure}[!t]
\centering
\includegraphics[width=8.5 cm]{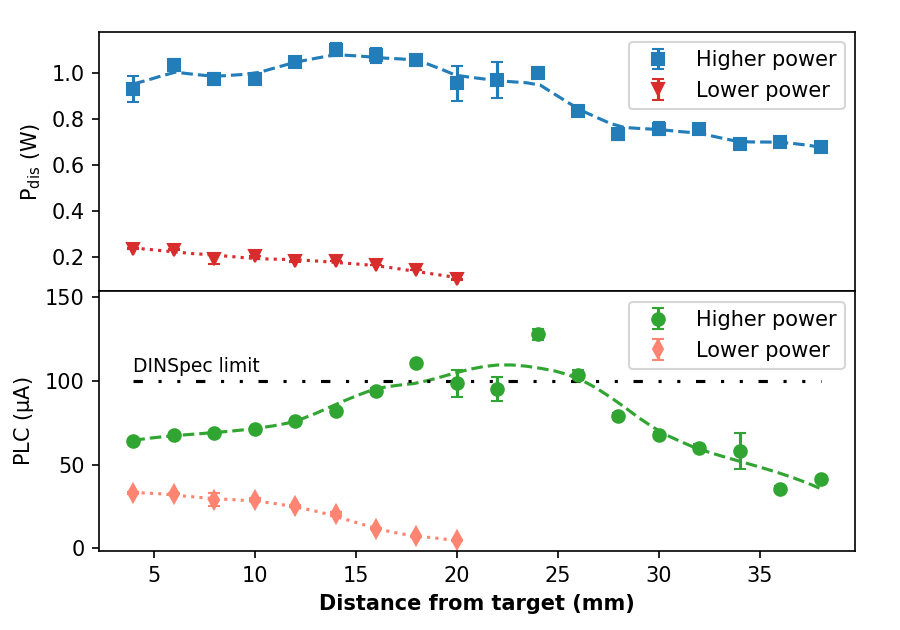}
\caption{Results for average discharge power and PLC as a function of the distance from the plasma outlet.\label{pwplc}}
\end{figure}

The results of $P_{dis}$ and PLC obtained for different $d$ values are presented in Fig.~\ref{pwplc}, for both higher and lower power operating conditions. As it can be seen in Fig.~\ref{pwplc}, the $P_{dis}$ values for the lower power curve decrease monotonically as $d$ is increased. However, for the higher power curve, the $P_{dis}$ values present an almost constant value for $d$ between 4 and 24 mm and start to decrease after that. A similar behavior was observed for the device presented in \cite{nastuta_cold_2022}.

The behavior of the PLC curves, in turn, are quite different in both cases. The lower power curve decreases monotonically while the higher power one presents a peak value almost at the middle of the measurement interval. Regarding the PLC values, for the lower power condition they are always below 50 {\textmu}A, which is less than half the DINSpec limit (100 {\textmu}A-AC). On the other hand, when operating in the higher power condition, the PLC values are above 100 {\textmu}A at some distances from the plasma outlet, reaching {$\sim$}130 {\textmu}A for $d$ = 24 mm. It is worth mentioning that both $P_{dis}$ and PLC values were measured operating with a network frequency of 50 Hz. Thus, when operating with a network frequency of 60 Hz it is expected to obtain $P_{dis}$ and PLC values 20{\%} and {$\sim$}10{\%} higher, respectively, than the ones presented in Fig.~\ref{pwplc}.

It is important to notice that the $\rm{PLC_{wf}}$ curves (see Fig.~\ref{wfEg}) can present peak values of the order of a few mA. However, in the DINSpec there is no mention of peak values in the $\rm{PLC_{wf}}$, and the PLC values lie in the {\textmu}A range.

\subsubsection{Gas temperature measurements}
\begin{figure}[!t]
\centering
\includegraphics[width=8.5 cm]{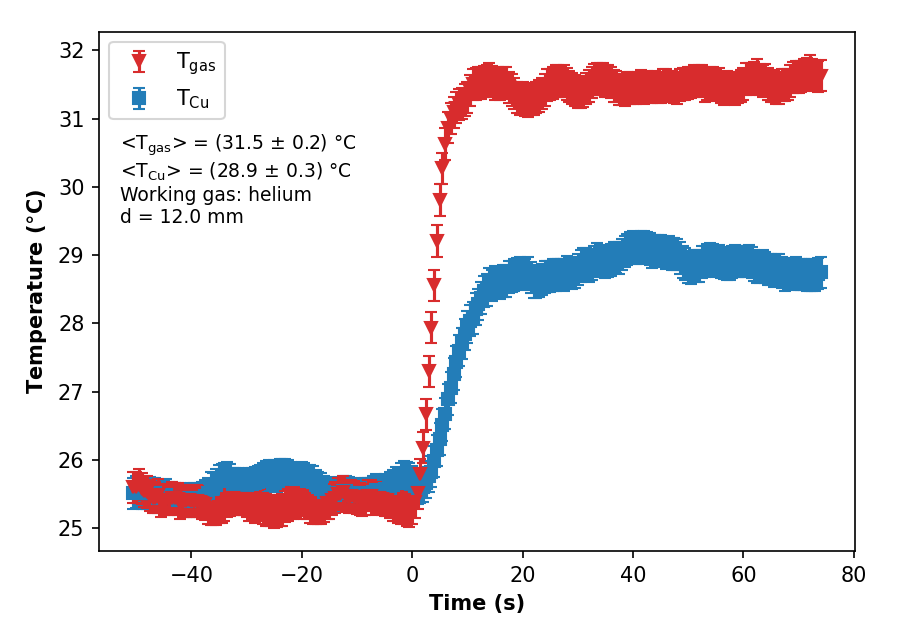}
\caption{Temporal evolution of $T_{gas}$ and $T_{Cu}$ for $d$ = 12.0 mm in the higher power condition.\label{tempTime}}
\end{figure}

Temperature measurements of gas and Cu plate ($T_{gas}$ and $T_{Cu}$, respectively) were performed simultaneously with temporal resolution by using the FOT sensors. Figure~\ref{tempTime} shows an example of the temporal evolution of both $T_{gas}$ and $T_{Cu}$ before the power supply be turned on (for t $<$ 0), and consequently without discharge ignition, and also after that (t $\geq$ 0). From that figure it can be seen that both $T_{gas}$ and $T_{Cu}$ tend to reach a steady state value after the plasma jet is turned on. Then, the temperature values for t $>$ 20 s are averaged to obtain the $\langle T_{gas} \rangle$ and $\langle T_{Cu} \rangle$ values for each distance ($d$) from the plasma outlet.

\begin{figure}[!ht]
\centering
\begin{subfigure}{0.48\textwidth}
\centering
\includegraphics[width=8.0 cm]{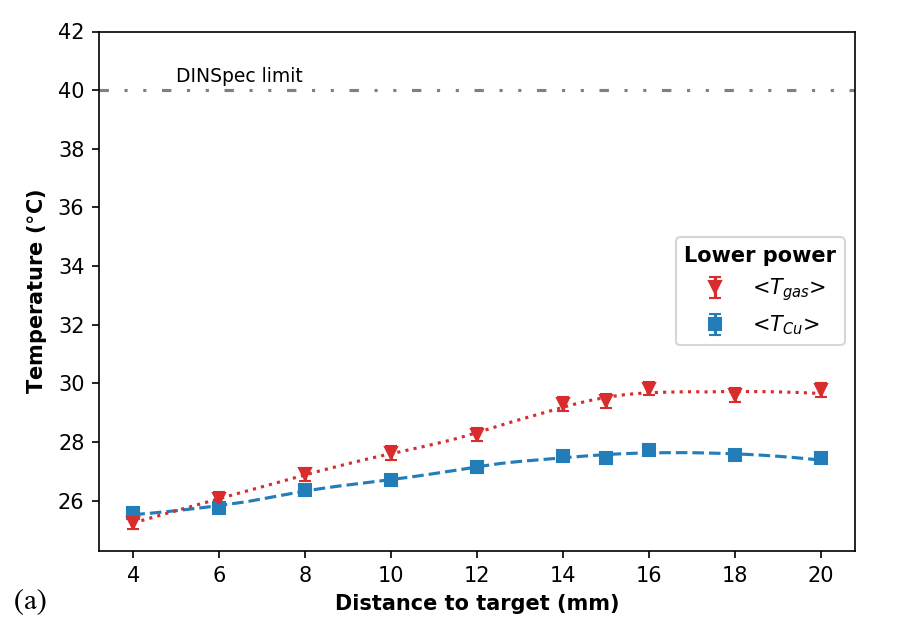}
\end{subfigure}

\begin{subfigure}{0.48\textwidth}
\centering
\includegraphics[width=8.0 cm]{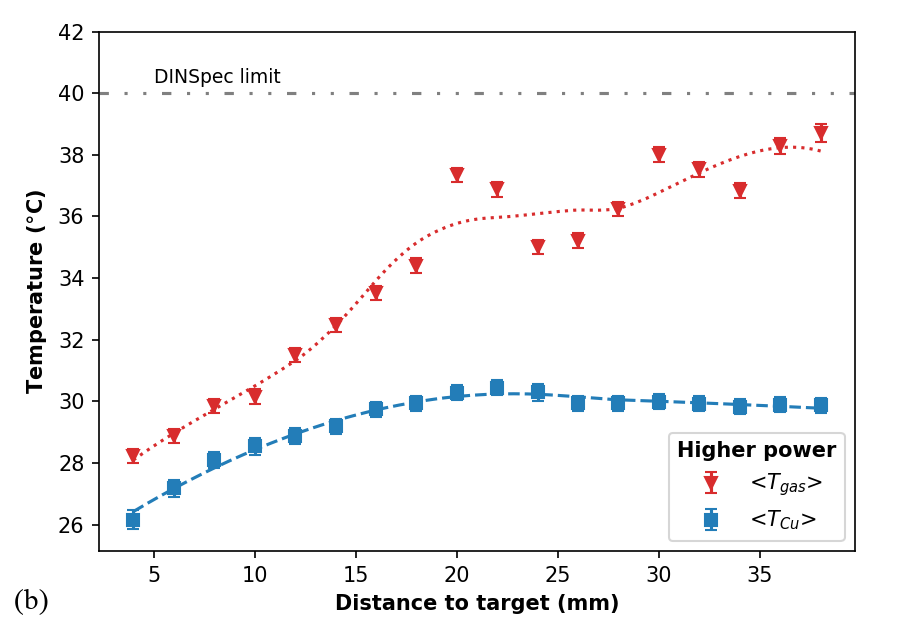}
\end{subfigure}

\begin{subfigure}{0.48\textwidth}
\centering
\includegraphics[width=8.0 cm]{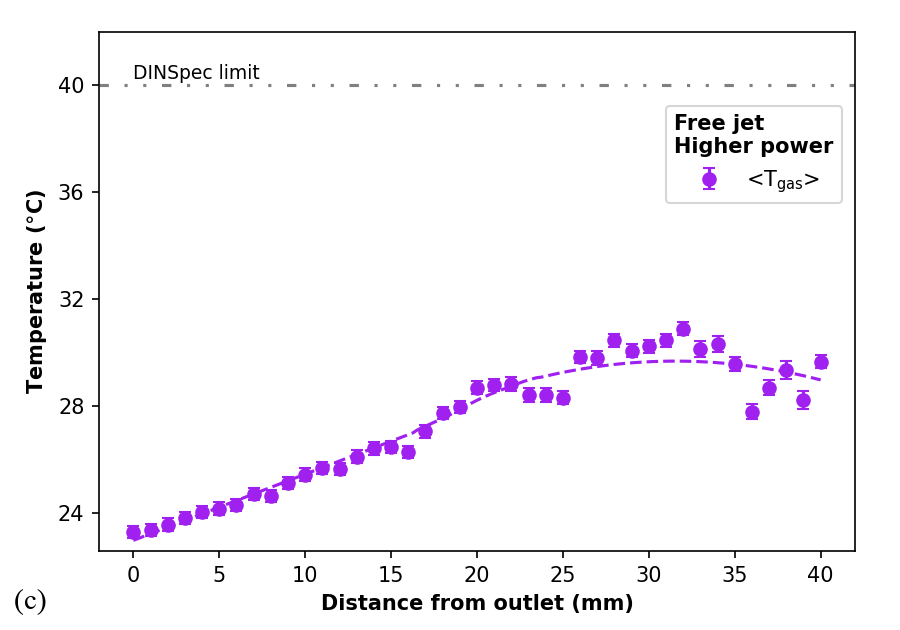}
\end{subfigure}

\caption{Average temperatures as a function of the distance from the plasma outlet: with the plasma jet impinging on the Cu plate for (a) the lower power, (b) the higher power conditions and (c) the free jet mode.}\label{temps1}

\end{figure}

The average values of the temperature measurements are presented in Fig.~\ref{temps1}, being that (a) and (b) show $\langle T_{gas} \rangle$ and $\langle T_{Cu} \rangle$ values as a function of $d$ for the plasma jet impinging on the copper plate for the lower and higher power conditions, respectively, while (c) shows the $\langle T_{gas} \rangle$ versus $d$ curve when the plasma jet is in the free mode, that is, without the copper target. As can be seen in Fig.~\ref{temps1}, the average gas temperature values of the plasma jets are below 40 {\textdegree}C in all cases and for the entire measurement ranges. However, from Fig.~\ref{temps1}(a,b) it is clear that both $\langle T_{gas} \rangle$ and $\langle T_{Cu} \rangle$ values increase as $d$ is incremented up to {$\sim$}15 mm in (a) and {$\sim$}20 mm in (b). For $d >$ 15 mm in (a), both temperature values tend to reach a plateau and for $d >$ 20 mm in (b), the $\langle T_{Cu} \rangle$ values also tend to stabilize but $\langle T_{gas} \rangle$ is still increasing. For the free jet condition (Fig.~\ref{temps1}(c)), the $\langle T_{gas} \rangle$ curve presents a behavior similar to the one measured with the Cu plate, but with lower values.

The oscillations observed in the $\langle T_{gas} \rangle$ curves presented in Fig.~\ref{temps1} are probably linked to a modulation of the gas flow when producing a plasma jet. This will be discussed in more detail in the image diagnostic section. Nevertheless, the growth trend observed in the $\langle T_{gas} \rangle$ values are directly related to the employment of helium as the working gas. In a recent work from our research groups, detailed measurements of the gas temperature as a function of the distance from the gas outlet revealed a consistent gas temperature growth as the distance from the gas outlet is incremented when flushing helium into the ambient air \cite{nascimento_gas_2023}. This is a phenomenon that happens even without discharge ignition and is one of the main heating sources of the gas in helium plasma jets with low values of dissipated power.

Despite the $\langle T_{gas} \rangle$ values being below the temperature limit suggested by the DINSpec, such an increase must be taken into account in applications that are sensitive to temperature variations.

\subsubsection{Evaluation of UV irradiation, production of RONS and $T_{vib}$ using OES}
\begin{figure}[!t]
\centering
\includegraphics[width=8.5 cm]{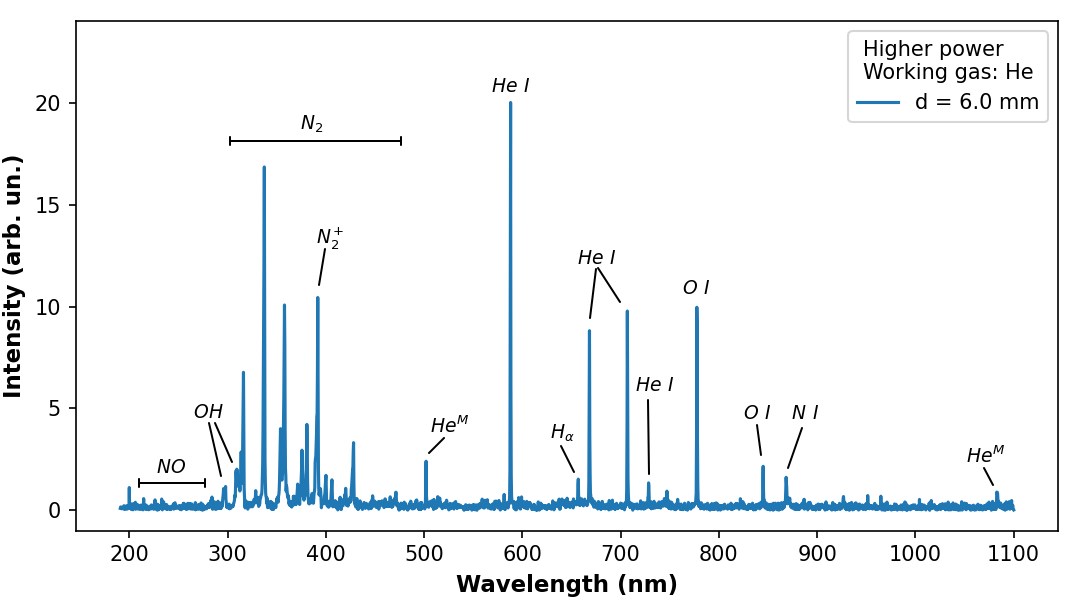}
\caption{Typical emission spectrum measured for the He plasma jet with the device operating in the higher voltage condition. The distance between the plasma outlet and the target was 6.0 mm.\label{sampSpec}}
\end{figure}

The identification of emitting species was aimed to obtain the most important reactive species produced within the plasma jet. A typical emission spectrum obtained for the plasma source operating in the higher power condition is shown in Fig.~\ref{sampSpec}. Molecular emissions from $\rm{OH}$ and $\rm{NO}$ were observed at 309 nm and in the wavelength range from 230-270 nm, respectively. An emission from the molecular nitrogen ion ($\rm{N_{2}^{+}}$) was observed at 391.4 nm. Atomic emissions from oxygen ($\rm{O}$, at 777.3, 844.6 and 926.6 nm) are also present in the spectra as well as atomic emissions from nitrogen ($\rm{N}$, 862$-$872 nm) and another one from hydrogen ($\rm{H}$, at 656.3 nm). Atomic emissions from He were also detected, being that two of them, at 501.6 nm and 1083 nm, are associated with transitions to metastable states.

\begin{figure}[!t]
\centering
\begin{subfigure}{0.48\textwidth}
\centering
\includegraphics[width=8. cm]{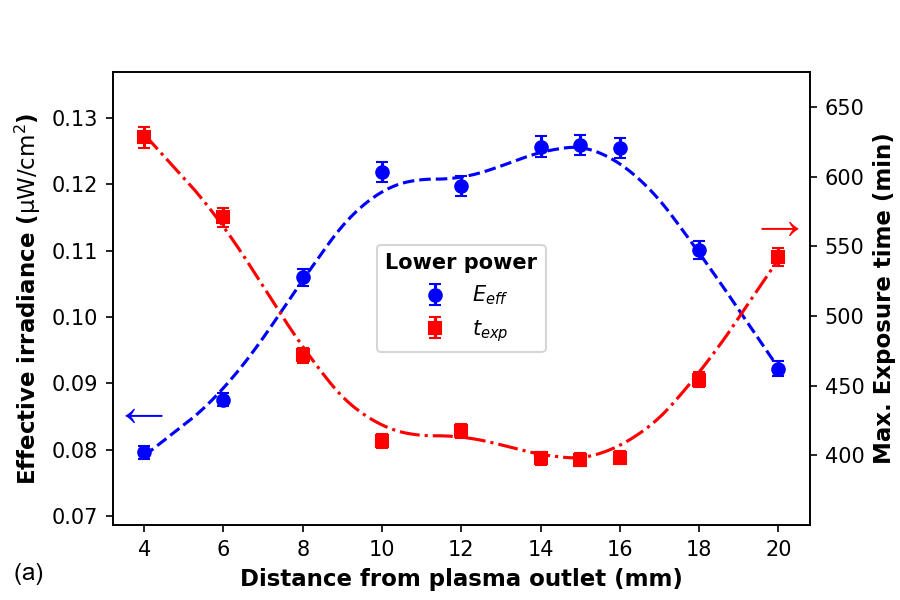}
\end{subfigure}
\begin{subfigure}{0.48\textwidth}
\centering
\includegraphics[width=8. cm]{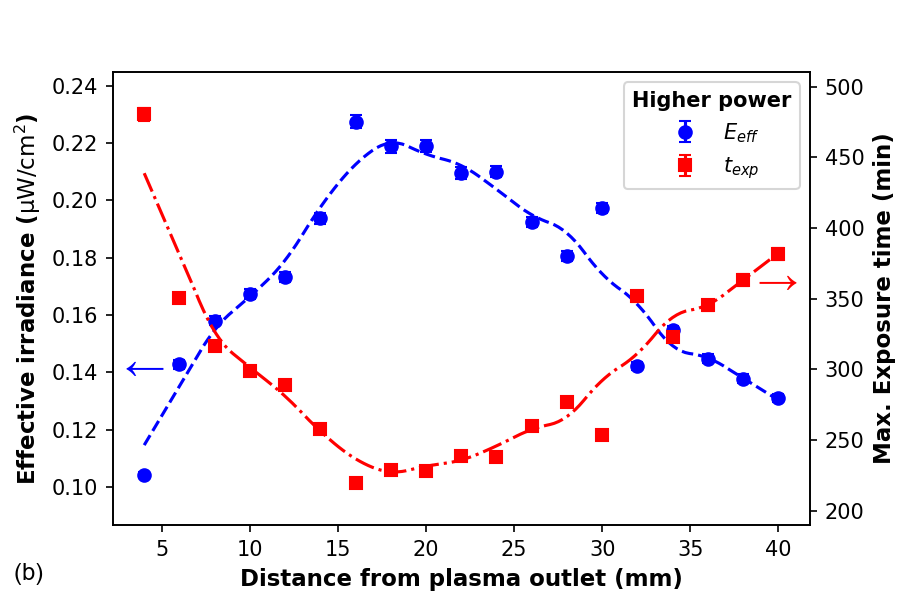}
\end{subfigure}
\caption{Effective irradiance and maximum exposure time as a function of the distance from the plasma outlet for the lower and higher power conditions (a and b, respectively).\label{IrrExp}}
\end{figure}

Figure~\ref{IrrExp} shows the results of both $E_{eff}$ and $t_{exp}$ as a function of the distance from the plasma outlet. The $t_{exp}$ values were calculated from the $E_{eff}$ ones using equation \ref{eqEeff}. From the curves shown in Fig.~\ref{IrrExp} it can be seen that the plasma jet produced with the portable device can be applied on human tissues for more than three hours in the worst case, which is in the higher power condition at $d$ = 16 mm. As it can be seen in Fig.~\ref{IrrExp}, the $E_{eff}$ values for the lower power case are, in general, much lower than the ones measured for the higher power condition, which leads to larger $t_{exp}$ values in the first case.

It can be also noticed that the $E_{eff}$ values do not change significantly for $d$ between 10 mm and 16 mm in the lower power case and also for $d$ between 16 mm and 24 mm in the higher power one. 

\begin{figure}[!t]
\centering
\begin{subfigure}{0.48\textwidth}
\centering
\includegraphics[width=8. cm]{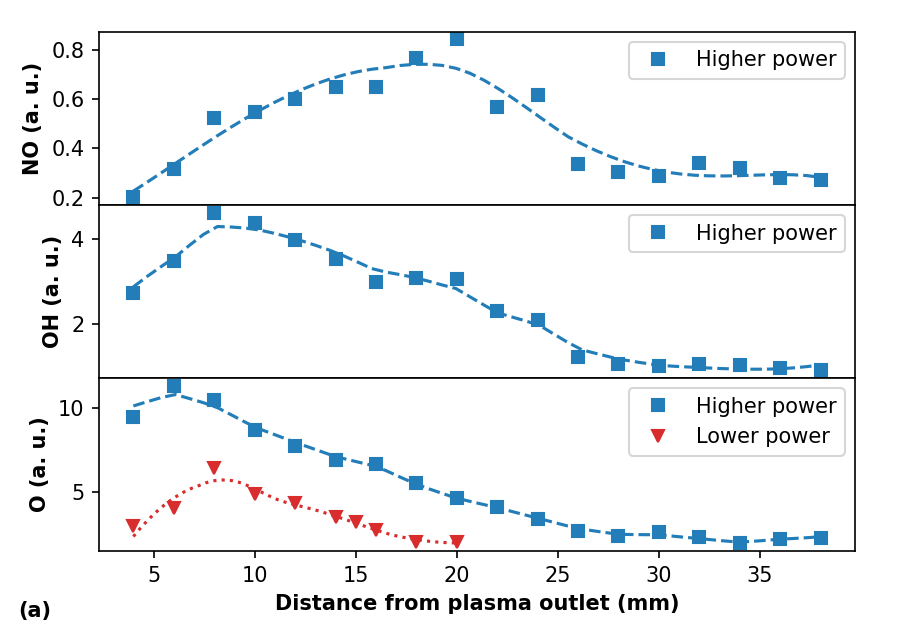}
\end{subfigure}
\begin{subfigure}{0.48\textwidth}
\centering
\includegraphics[width=8. cm]{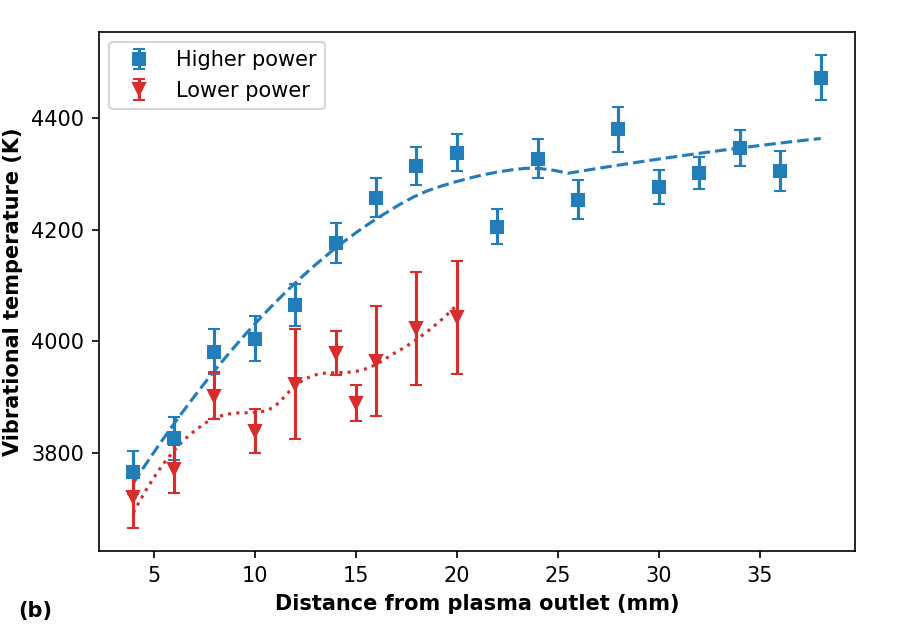}
\end{subfigure}
\caption{Curves of (a) intensity of light emitted by different reactive species in excited states and (b) Variation of the vibrational temperature of $\rm{N_{2}}$ molecules, both as a function of $d$.\label{rSpTv}}
\end{figure}

Figure~\ref{rSpTv} shows curves of (a) the intensity of the light emitted by some important reactive species present in the plasma jet in excited states $\rm{NO}$ (at 236 nm), $\rm{OH}$ (at 308 nm) and $\rm{O}$ (at 777 nm) and (b) the vibrational temperature ($T_{vib}$) values of the $\rm{N_2}$ molecules, both as a function of the distance from the plasma outlet. The intensity of the light emitted by excited species in the plasma jet is proportional to its abundance. As it can be seen in Fig.~\ref{rSpTv}(a), the amount of excited $\rm{NO}$ produced by the plasma jet tends to increase in the $d$ interval from 4 mm to 20 mm, presenting a reduction trend after that. Regarding the excited $\rm{OH}$ radical, it also presents a growth trend close to the plasma outlet, peaking at $d$ {$\approx$} 8 mm, followed by a reduction trend. The intensity of the light emitted by the excited $\rm{NO}$ and $\rm{OH}$ radicals in the lower power condition are very low, with a low signal to noise ratio, so in this case the production of such species is not significant compared to the higher power condition. In relation to the production of atomic oxygen, in both conditions (higher and lower power) there is a growth trend close to the plasma outlet, followed by a downward trend after that.

The $T_{vib}$ values measured for the $\rm{N_{2}}$ molecules in the plasma jet present a growth trend in the entire measurement range. The vibrational temperature of molecules in a gas reveals their reactivity, that is, molecules at higher $T_{vib}$ values are more likely to chemically react with a substrate \cite{lambert_vibrationvibration_1967, smith_preference_2004}. The vibrational temperature values of $\rm{NO}$ and $\rm{OH}$ may be different from those obtained for $\rm{N_{2}}$.

It is important to notice that the curves presented in Fig.~\ref{rSpTv}(a) does not represent the axial distribution of the species in the plasma jet, since such OES measurements were performed using the scheme shown in Fig.~\ref{expSetPlSrc}(b) and the intensity emission spectra is then integrated along the entire path between the plasma outlet and target. It is worth mentioning that the $T_{vib}$ values obtained in this way are the average ones for this parameter in the entire plasma jet at each $d$ value.

\subsubsection{Concentration of $\rm{O_3}$ and $\rm{NO_x}$ produced within the plasma jets}
\begin{figure}[!ht]
\centering
\begin{subfigure}{0.4\textwidth}
\centering
\includegraphics[width=6 cm]{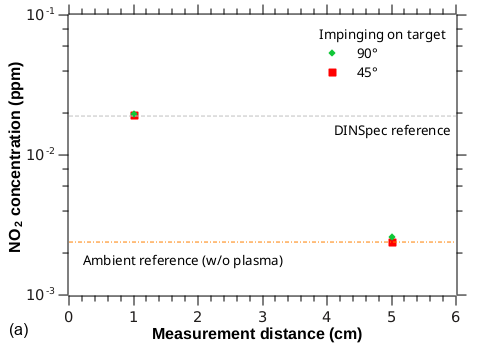}
\end{subfigure}

\begin{subfigure}{0.4\textwidth}
\centering
\includegraphics[width=6 cm]{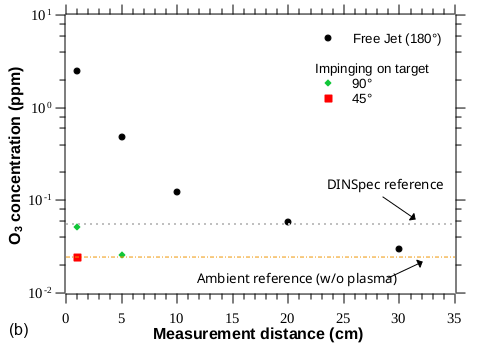}
\end{subfigure}

\begin{subfigure}{0.4\textwidth}
\centering
\includegraphics[width=6 cm]{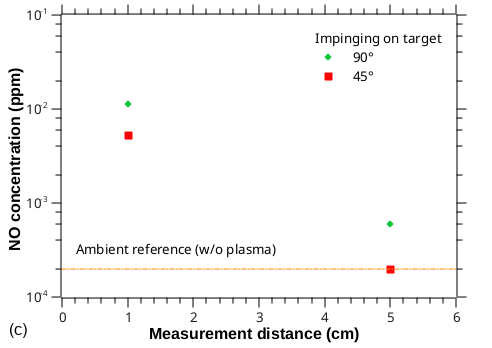}
\end{subfigure}
\caption{Concentration of harmful gases as a function of the distance from the plasma source: (a) $\rm{NO_2}$; (b) $\rm{O_3}$; (c) $\rm{NO}$. The distance between the plasma outlet and target was 15.0 mm.}\label{O3NOx}

\end{figure}

The results of $\rm{O_3}$ and NO/$\rm{NO_2}$ concentrations measured as a function of the distance from the plasma jet ($D$) are presented in Fig.~\ref{O3NOx}.

For those measurements, the distance $d$ between the plasma outlet and the target was set to 15.0 mm. In Fig.~\ref{O3NOx} are also displayed the DINSpec reference values for the concentrations of $\rm{NO_2}$ and $\rm{O_3}$ as well as the $\rm{O_3}$, $\rm{NO}$ and $\rm{NO_2}$ concentrations in the ambient air when the plasma source is powered off. Such measurements were performed only for the higher power operating condition, since a preliminary test revealed that for the lower power case the concentration of those gasses had no significant difference from the reading values without plasma ignition. Each data point in Fig.~\ref{O3NOx} is the average of ten measurements, taken once every 30 seconds, with the plasma jet in continuous operation. From the curves shown in Fig.~\ref{O3NOx}, it can be seen that the emissions of both $\rm{O_3}$ and $\rm{NO_2}$ are below the limit established by the DINSpec when the plasma jet impinges on the Cu plate (for both measurements performed at 45{\textdegree} and at 90{\textdegree}). For all the measured gasses, their concentrations decrease as a function of $D$ up to their respective values found without ignition of the plasma jet. Figure~\ref{O3NOx}(b) also shows the results obtained for the $\rm{O_3}$ production as a function of $D$ for the free jet operation (at 180{\textdegree}, without the Cu target). In this case, an $\rm{O_3}$ concentration much higher than the recommended value for $D <$ 20 cm is observed. However, the $\rm{O_3}$ concentration tends to decrease as $D$ is increased.

Based on the information obtained in these experiments, it can be said that the $\rm{NO_2}$ and $\rm{O_3}$ concentrations should not be an issue when operating with the plasma jet impinging on a target, which could be a part of a patient's body. However, the operation of the plasma source in the free jet mode must be avoided, limited to the shortest possible time interval or supported by a suction system removing the toxic gas.

\subsubsection{Image diagnostics}
\begin{figure*}[!ht]
\centering
\begin{subfigure}{0.48\textwidth}
\centering
\includegraphics[width=8.0 cm]{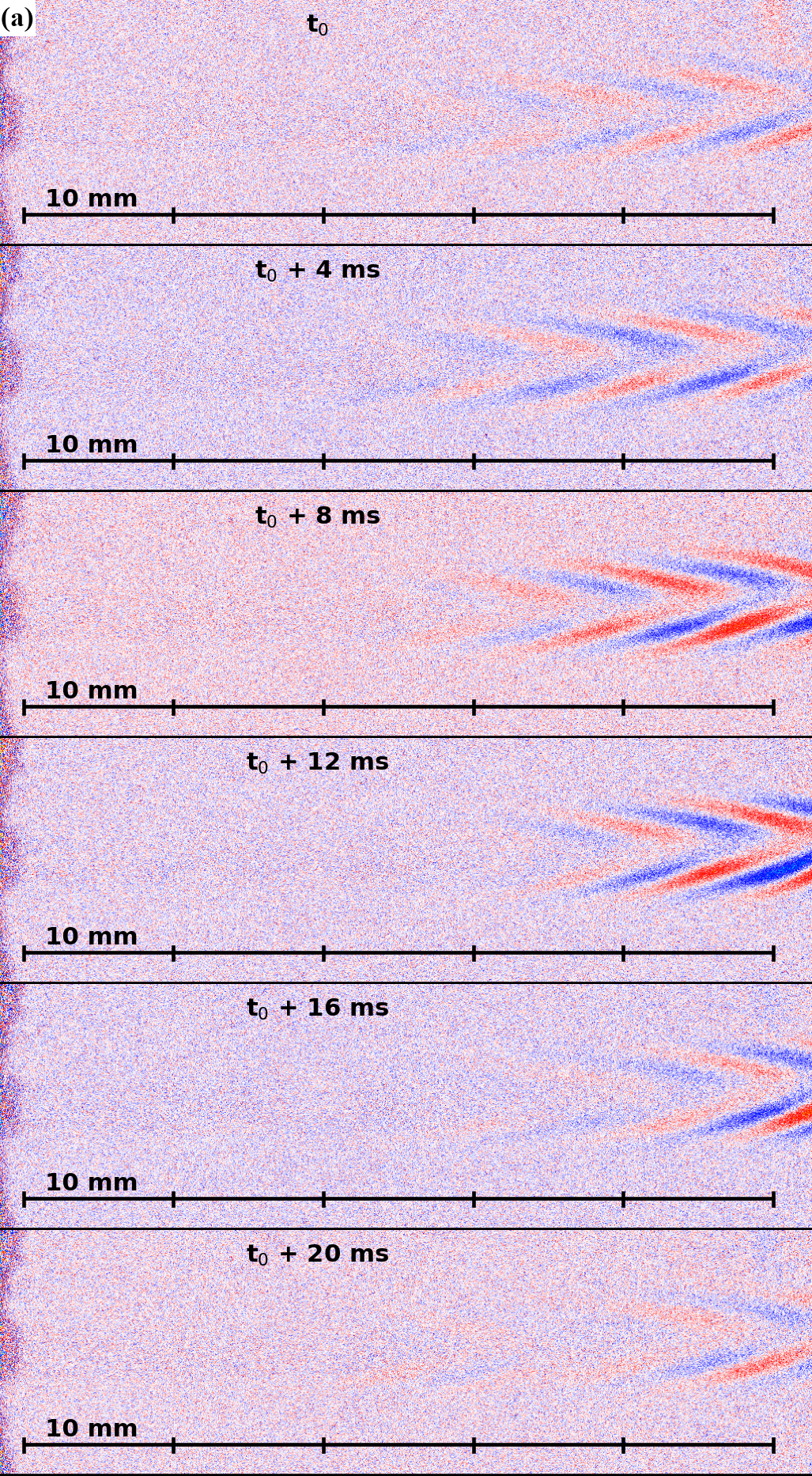}
\end{subfigure}
\begin{subfigure}{0.48\textwidth}
\centering
\includegraphics[width=8.0 cm]{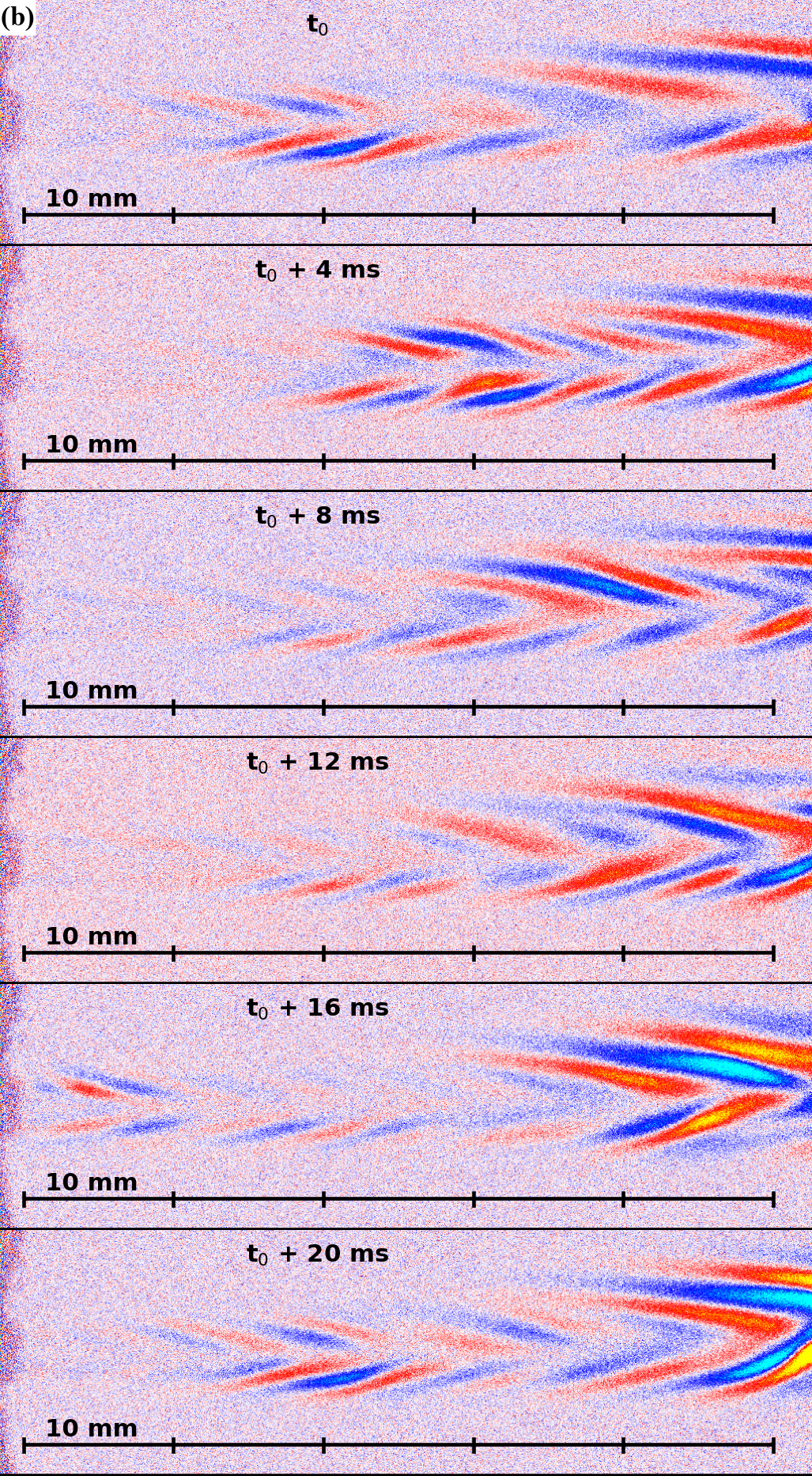}
\end{subfigure}
\caption{Background subtracted Schlieren images of the He gas flow with the discharge (a) off and (b) on for different time instants. The tube outlet is aligned with the beginning of the ruler, on the left side of the images. Gas flows from left to right. Blue parts indicate negative, red parts positive deviation from the mean image.\label{schImages}}
\end{figure*}

In our study, we expanded upon the standard characterizations as suggested by DINSpec, incorporating the Schlieren imaging technique to analyze the helium gas flow in the free jet and free flow modes under conditions with and without discharge ignition, respectively. Figure~\ref{schImages} depicts representative Schlieren images for (a) plasma-off and (b) plasma-on states.

\begin{figure}[ht]
\centering
\begin{subfigure}{0.55\textwidth}
\centering
\includegraphics[width=1\textwidth]{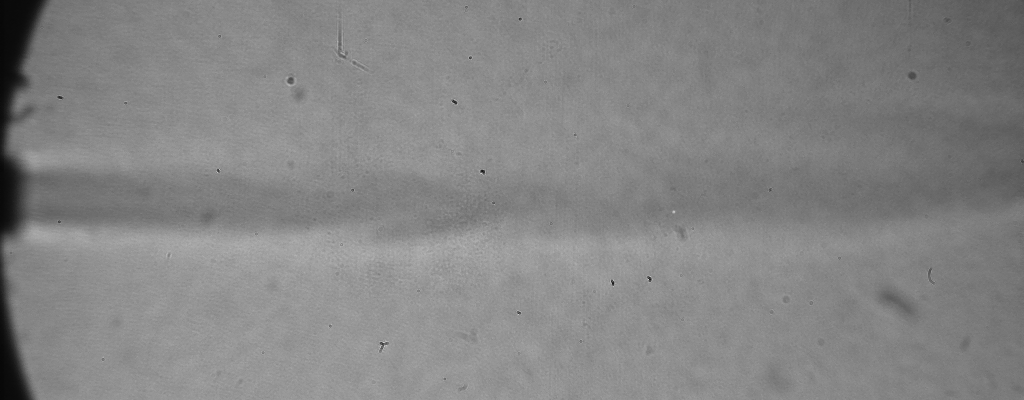}
\end{subfigure}
\begin{subfigure}{0.55\textwidth}
\centering
\includegraphics[width=1\textwidth]{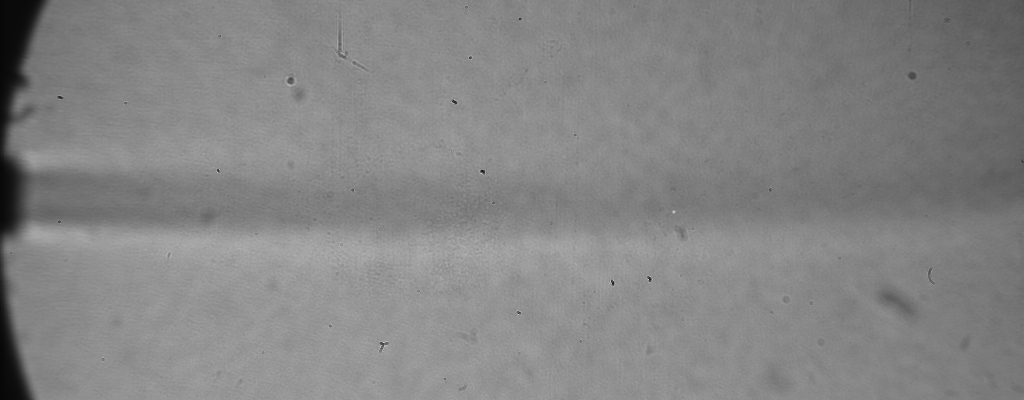}
\end{subfigure}
\begin{subfigure}{0.55\textwidth}
\centering
\includegraphics[width=1\textwidth]{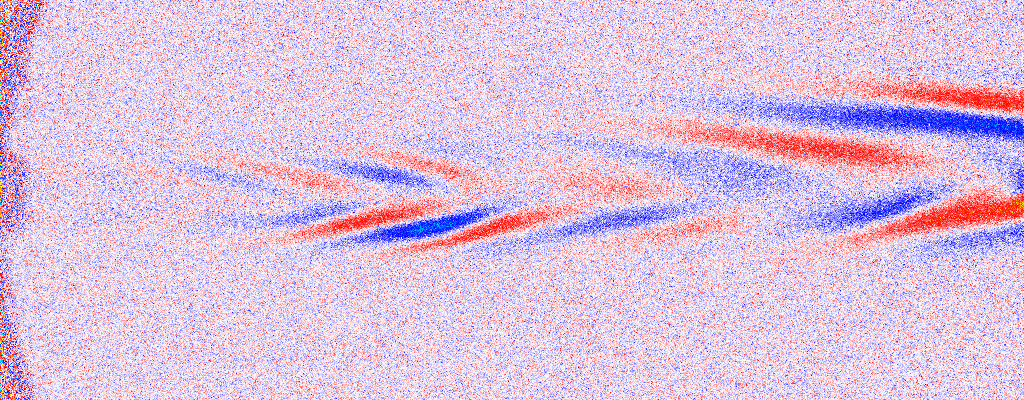}
\end{subfigure}
\caption{Image processing example. Original Schlieren image (top), mean image (center), and resultant subtracted image with color gradient (bottom).\label{imgTreat}}
\end{figure}

To delineate transient deviations in flow structures from the mean behavior, we implemented a normalization technique wherein each frame within a series was subtracted by the temporal mean of the entire series. This computational approach accentuates features in individual frames that deviate from the collective mean flow characteristics observed throughout the duration. Figure~\ref{imgTreat} provides a visual representation of the method: the original Schlieren image is displayed on the top, the mean image in the center, and the resultant subtracted image at the bottom. Gas emanates from the tube outlet on the left and progresses rightward. Owing to density disparities and consequent refractive index variations, helium, being less dense, is represented darker, while the ambient air appears brighter. In the subtracted images of Fig.~\ref{schImages}, the hues of blue and red denote pronounced shifts in optical density.

In the absence of plasma, the flow remains homogeneous for an initial span post-exit (from 0 to approximately 20 mm). This uniformity is discernible in the subtracted images, evident from the noise-like pattern, signifying minimal deviation from the mean. However, beyond the 20 mm mark, the homogeneity dissipates, giving rise to an alternating, arrow-shaped pattern that traverses in the streamwise direction. This is indicative of the incipient mixing phase between helium and ambient air. Factors such as volumetric flow rate, nozzle diameter, and gas density ratio likely influence this breakup length. Conversely, in the presence of plasma, there is an immediate disruption in flow homogeneity right after the outlet, leading to a pattern akin to the aforementioned mixing sequence.

Both flow conditions $-$plasma-on and plasma-off$-$ demonstrate a mild inclination to veer upwards, suggesting potential buoyancy forces at play. Additionally, they exhibit a periodic flow-wise modulation pattern over time. A comparative analysis of images at times $\rm{t_0}$ and $\rm{t_0 +}$ 20 ms for both flow states reveals analogous intensities spanning the jet length, potentially alluding to a stable, stream-wise modulation intrinsically linked to volumetric flow rate, nozzle diameter, and gas density ratio as well.

\subsection{Biological assays}
\subsubsection{Inhibitory effect of CAPP on different microorganisms}
\begin{figure}[t]
\centering
\begin{subfigure}{0.48\textwidth}
\includegraphics[width=1\linewidth]{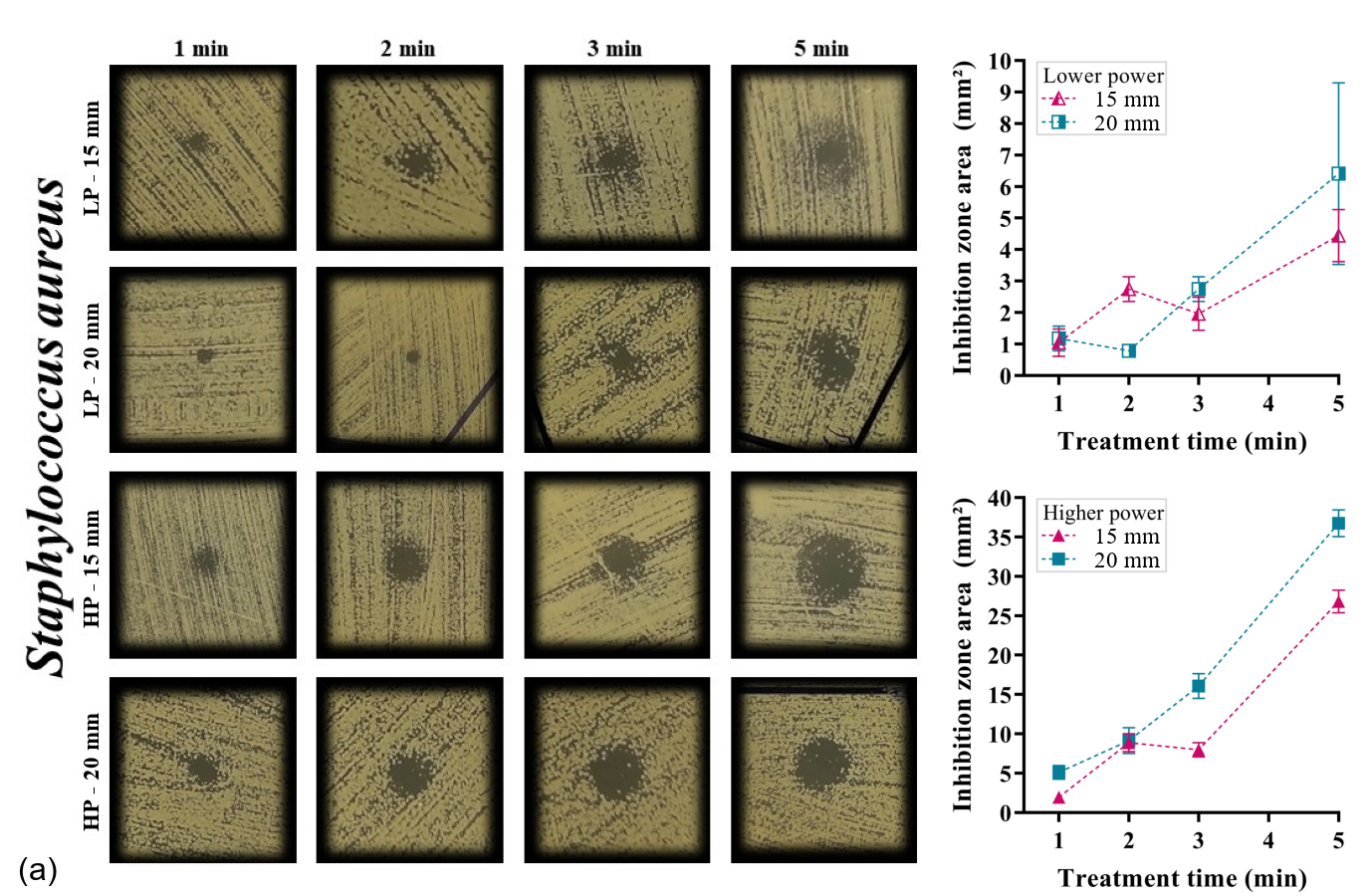}
\end{subfigure}
\begin{subfigure}{0.48\textwidth}
\includegraphics[width=1\linewidth]{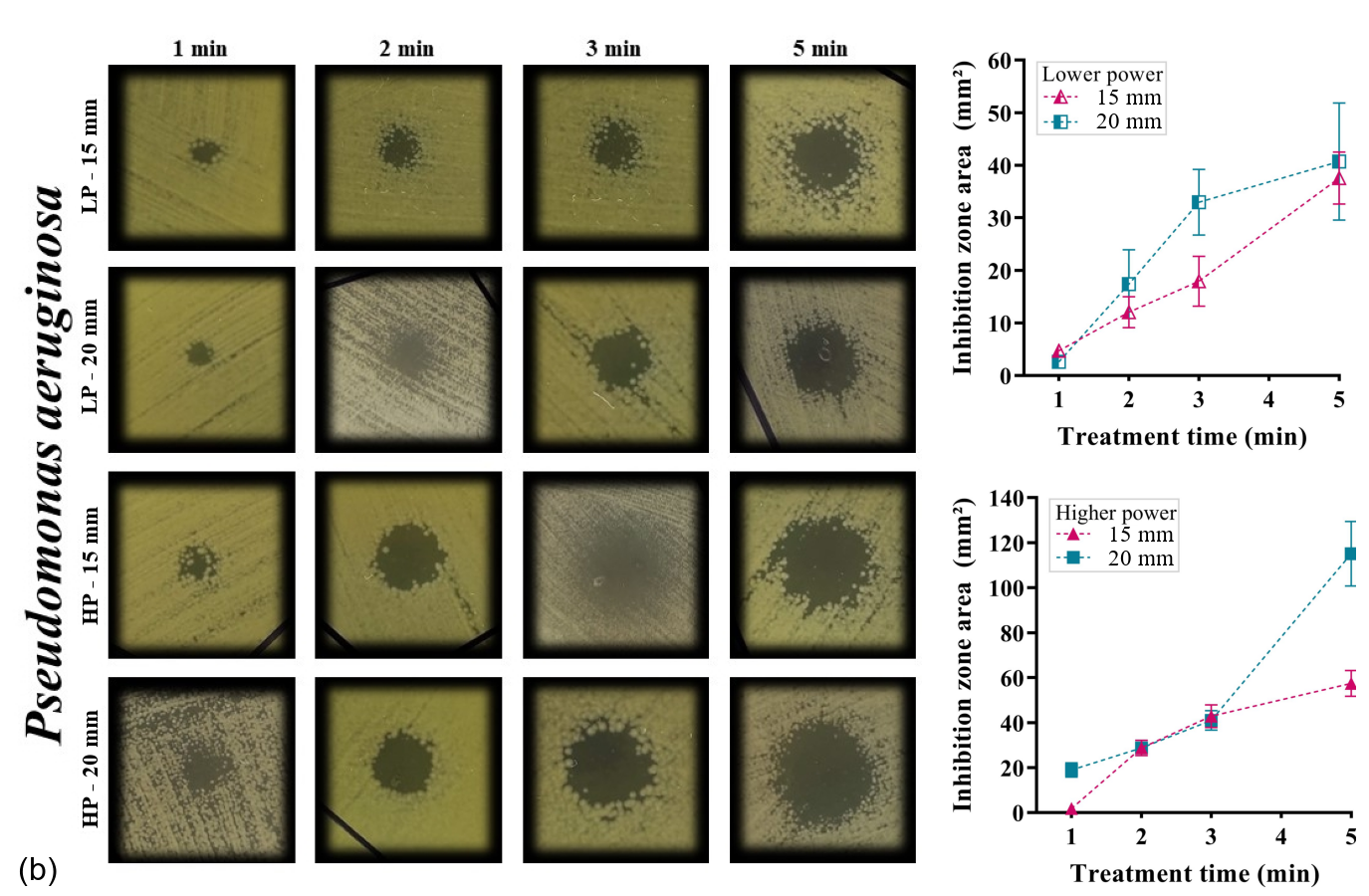}
\end{subfigure}
\begin{subfigure}{0.48\textwidth}
\includegraphics[width=1\linewidth]{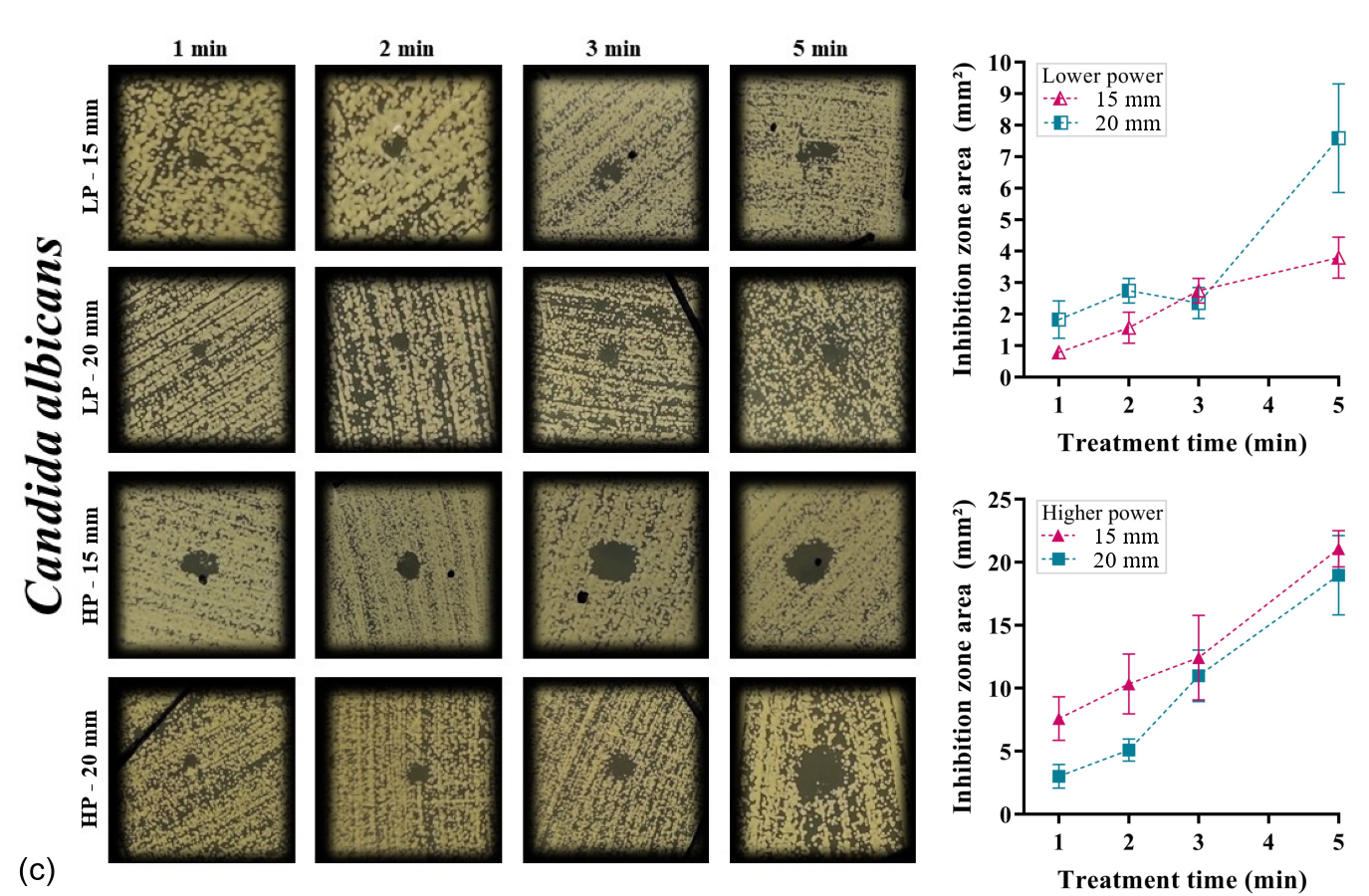}
\end{subfigure}
\caption{A schematic representation of the area zone of inhibition as well as a line graph illustrating the mean and standard deviation. Antimicrobial efficacy of He plasma jet against bacterial pathogenic strains (a) \textit{Staphylococcus aureus}, (b) \textit{Pseudomonas aeruginosa} and (c) fungal species \textit{Candida albicans} at different exposure times, distance from plasma outlet and discharge power values.\label{inhZone}}
\end{figure}

The atmospheric plasma of helium gas is capable of generating several chemical reactions when interacting with air, forming reactive oxygen and nitrogen species (ROS and RNS) and show biological effects \cite{lu_roles_2008, lotfy_inactivation_2020}. In the present work, the results on Fig.~\ref{inhZone}(a-c) highlights a positive correlation between the exposure period, distance, and discharge power with the area of microbial inhibition. Furthermore, it was detected that the microorganisms showed different susceptibilities to the plasma jet. 

Increasing the distance from 15 to 20 mm in the higher power condition of the APPJ device leads to an increase in antibacterial efficacy over time of exposure. For \textit{S. aureus}, it was detected that after 1 min of exposure, the inhibition area increased from 1.96 {\textpm} 1.28 mm${}^2$ to 5.10 {\textpm} 2.14 mm${}^2$ and after 5 min of exposure from 26.82 {\textpm} 3.52 mm${}^2$ to 36.76 {\textpm} 4.16 mm${}^2$, respectively. For \textit{P. aeruginosa}, higher values of inhibition halos of 1.83 {\textpm} 1.46 mm${}^2$ to 18.97 {\textpm} 7.69 mm${}^2$ could be be observed after 1 min exposure and from 57.43 {\textpm} 14.02 mm${}^2$ to 115.13 {\textpm} 35.34 mm${}^2$ after 5 min. However, a slight decline in the zone of inhibition was observed for \textit{C. albicans} when the distance was increased. After 1 min of exposure, values from 7.58 {\textpm} 4.23 mm${}^2$ to 3.00 {\textpm} 2.29 mm${}^2$ and, after 5 min of exposure, from 21.06 {\textpm} 3.52 mm${}^2$ to 18.97 {\textpm} 7.69 mm${}^2$ were observed. In general, from the perspective of a single parameter with antibacterial and antifungal action, high antimicrobial efficacy was observed in the high power condition, at a distance of 15 mm and a treatment time of 5 min. The differences detected among the species can be attributed to the diversity in cell structure and composition. Gram-positive bacteria stand out for having a thicker peptidoglycan layer compared to Gram-negative bacterial species, playing an important protective role \cite{silhavy_bacterial_2010}. An investigation performed by Yusupov \textit{et al}, revealed that ROS such as $\rm{O_3}$, $\rm{O_2}$, and $\rm{O}$ atoms generated by non-thermal plasma are responsible for breaking important bonds of the peptidoglycan structure ($\rm{C-O}$, $\rm{C-N}$, and $\rm{C-C}$), a factor that can contribute to bacterial destruction \cite{yusupov_atomic-scale_2012}.

When working under the higher power condition and exposing the microorganisms to plasma treatment at different distances from the plasma outlet, the power dissipated in the plasma jet remained almost constant, around 295 mW. However, from the gas temperature results shown in Fig.~\ref{temps1}, the $\rm{NO}$ production shown in Fig.~\ref{rSpTv}(a) and the vibrational temperature values presented in Fig.~\ref{rSpTv}(b), it can be inferred that the better results shown in Fig.~\ref{inhZone} (a and b) obtained at 20 mm from the plasma outlet are probably linked to a synergy among higher $T_{gas}$ and $T_{vib}$ values together with a higher $\rm{NO}$ production when compared to the values obtained for such parameters for $d$ close to 15 mm. According to the data shown in Fig.~\ref{IrrExp}, a possible contribution of the UV irradiation when working at different $d$ values may not be expressive. The spread of reactive species produced by APPJ may also be contributing to the increase in the effective area of the inhibition zones. The reactive species could spread more widely over distance, as indicated by the enhanced flow cross section detected with Schlieren imaging, and would therefore increase the area of the inhibition zone, on average, while the center becomes saturated, the outer areas become more exposed to the reactive species over time.

\subsubsection{Evaluation of CAPP cytotoxicity}

\begin{figure}[!t]
\centering
\begin{subfigure}{0.24\textwidth}
\centering
\includegraphics[width=4.2 cm]{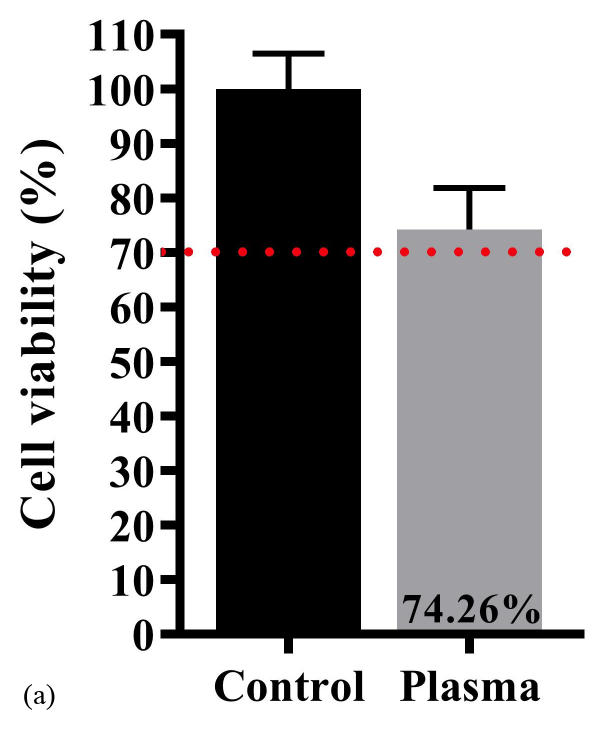}
\end{subfigure}
\begin{subfigure}{0.24\textwidth}
\centering
\includegraphics[width=4.2 cm]{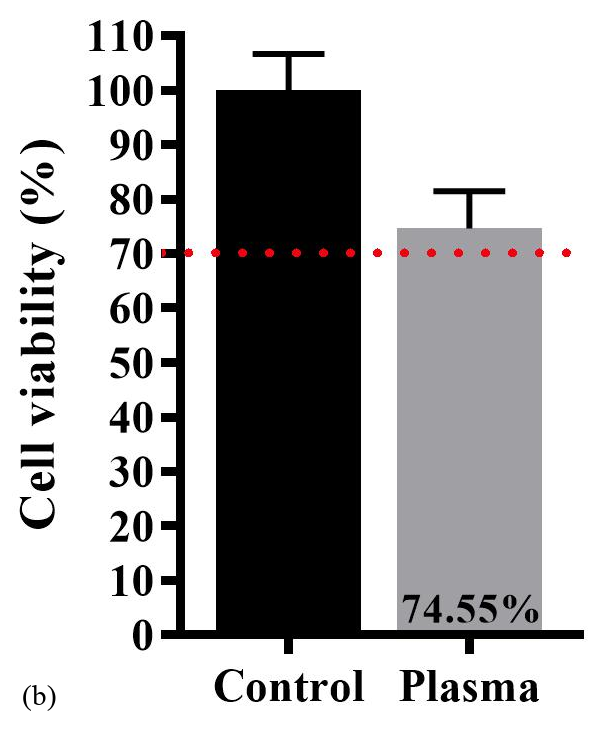}
\end{subfigure}
\caption{Viable cells obtained with MTT assay for (a) Keratinocytes and (b) fibroblasts. The dashed line indicates the normative value of cell viability (70{\%}).}\label{cellTox}
\end{figure}

In Fig.~\ref{cellTox} are presented the results of cytotoxicity assessment carried out for two cell types: (a) keratinocytes and (b) fibroblasts. The conditions chosen for these assays were with the device operating at the high power condition, with the sample placed at 15 mm from the plasma outlet and 5 min of plasma exposure. This choice was made because these conditions presented the most effective antimicrobial activity for the fungus \textit{C. albicans}, which is the most resistant microorganism among those tested in this work. From Fig.~\ref{cellTox} it can be seen that the plasma treatment reduced the cell viability to 74.26{\%} for keratinocytes and 74.55{\%} for fibroblasts. In spite of this, the results are within the normative range of cell viability, which suggests a value $\geq$70{\%} \cite{international_organization_for_standardization_iso_2009}.

The results obtained for cell viability tested on two different cell types reinforce the fact that the plasma jet parameters ($P_{dis}$, PLC and $T_{gas}$), as well as the production of harmful gasses and UV radiation, are at safe levels for both the conditions tested and the exposure time adopted for the tests.

\section{Conclusions}
In this work a low cost and low repetition rate plasma source aimed for medical and biomedical applications was characterized according to the DINSpec protocol. As a general conclusion, the device presented good antimicrobial efficacy and low cytotoxic effects on healthy cells. The antimicrobial efficacy can be attributed to the overall set of discharge parameters, that is, discharge power, electrical current, gas temperature and production of reactive species, which, together, generate the necessary conditions to interact with the microorganisms exposed to plasma treatment in order to inhibit their development. The low cellular cytotoxicity presented by the plasma jet produced with such a device is a good indication that the discharge parameters are in conformity with the safety requirements for medical applications. Furthermore, by the characterization results of the plasma source, it can be said that the device is capable of operating within the safety standards established by the DINSpec.

Regarding the PLC results, when operating using the higher power condition the PLC values measured at certain distances were above the limit suggested by the DINSpec. Since this is a plasma source that is still under development, it can be adjusted in order to prevent such a situation.

Regarding the gas temperature values obtained for the plasma jet, the results show that they are always below the 40 {\textdegree}C threshold established for medical applications in any operating conditions studied in this work. However, due to the use of helium as the working gas, the $T_{gas}$ values presented a growth trend as the distance between the plasma outlet and target is increased. The curves of $T_{gas}$ as a function of $d$ presented oscillations as $d$ was increased for the higher power condition. Such oscillations in the $T_{gas}$ values may be related to the modulation of the gas flow observed through Schlieren measurements.

The amount of UV irradiation emitted by the plasma source is considerably low, which allows plasma treatment using long exposure times. It was also found that the concentration of gasses that are potentially nocive for human tissues are below the limits recommended by the DINSpec.

As a general conclusion, it can be said that the device under study in this work complies with the safety requirements defined by the DINSpec. Of course, some fine adjustments will be required in order to obtain a commercial prototype. Nevertheless, the overall results can be considered good for a low cost device. Although the device uses helium as the working gas, which has high acquisition costs, it has been demonstrated that the antimicrobial efficacy is quite satisfactory. Thus, since most treatments must be of short duration, the operation tends to have a good cost-benefit ratio.

It is interesting to note that, when exposing the \textit{S. aureus} and \textit{P. aeruginosa} bacteria to APPJ treatment, the best results were obtained for a larger distance between the plasma outlet and the substrate. However, the same did not happen for \textit{C. albicans} exposed to plasma treatment. The first result may indicate that the higher values of total production of $\rm{NO}$ together with the higher $T_{gas}$ and $T_{vib}$ values at larger distances obtained for the APPJ, as well as the enhanced spreading of species as indicated by the enhanced flow cross section detected with Schlieren imaging, produce the better conditions for bacteria inactivation.

In future works the studies will be focused on the use of working gasses other than helium for the production of plasma jets. The main alternative will be to work with argon, trying to find a suitable admixture that avoids the filamentary discharges produced when this is the working gas.

\section*{Acknowledgments}
This work received financial support from the São Paulo Research Foundation$-$FAPESP under grants 2019/05856-7, 2019/25652-7, 2020/09481-5, 2021/00046-7 and 2021/14391-8.

%
%



\bibliographystyle{ieeetr}
\bibliography{portable_device_paper}

\end{document}